\documentclass[a4paper,aps,pra,10pt,twocolumn,floatfix,superscriptaddress,notitlepage,usenames,dvipsnames,svgnames,table,nofootinbib,longbibliography]{revtex4-2}
\pdfoutput=1
\usepackage[utf8]{inputenc}
\usepackage[T1]{fontenc}
\usepackage{graphicx}
\usepackage{grffile}
\usepackage{wrapfig}
\usepackage{rotating}
\usepackage[normalem]{ulem}
\usepackage{amsmath}
\usepackage{textcomp}
\usepackage{amssymb}
\usepackage{capt-of}
\usepackage{soul}
\usepackage{parskip}
\usepackage{mathrsfs}
\usepackage[margin= 0.75in]{geometry}
\usepackage[braket, qm]{qcircuit}
\usepackage{footmisc}

\usepackage{pgfplots}
\definecolor{blue-violet}{rgb}{0.54, 0.17, 0.89}

\usepackage[english]{babel}
\usepackage{letltxmacro}
\LetLtxMacro{\ORIGselectlanguage}{\selectlanguage}
\makeatletter
\DeclareRobustCommand{\selectlanguage}[1]{%
  \@ifundefined{alias@\string#1}
    {\ORIGselectlanguage{#1}}
    {\begingroup\edef\x{\endgroup
      \noexpand\ORIGselectlanguage{\@nameuse{alias@#1}}}\x}%
}
\newcommand{\definelanguagealias}[2]{%
  \@namedef{alias@#1}{#2}%
}
\makeatother
\definelanguagealias{en}{english}
\definelanguagealias{eng}{english}

\usepackage{bbm}
\usepackage{etoolbox}
\usepackage{tcolorbox} 
\usepackage{tikz}
\usetikzlibrary{patterns}
\usepackage{amsthm}
\usepackage{amssymb}
\usetikzlibrary{arrows.meta}
\usepackage{mathtools}

\usepackage{bm}
\usepackage{dsfont}
\usepackage[font=small,labelfont=bf,justification=RaggedRight,format=plain]{caption}
\usepackage{subcaption}
\usepackage{enumerate}
\usepackage{hhline}

\usepackage{thmtools}
\usepackage{thm-restate}

\usepackage{microtype}
\usepackage{dsfont}
\usepackage{tensor}

\usepackage{physics}
\usepackage{mathrsfs}
\usepackage{mathtools}
\usepackage{fixltx2e}
\usepackage{relsize}

\DeclarePairedDelimiterX\phys[2]{\langle}{\rangle}{#1 \delimsize\vert\mathopen{} #2}

\theoremstyle{remark}

\usepackage{tikz}
\usepackage{tikz-network}
\usetikzlibrary{patterns,decorations.pathreplacing,decorations.pathmorphing}

\usepackage{color}
\definecolor{myred}{RGB}{232,102,102}
\definecolor{myblue}{RGB}{187,187,255}
\definecolor{myorange0}{RGB}{252,226,5}
\definecolor{myorange0c}{RGB}{255,255,255}
\definecolor{myorange}{RGB}{255,165,0}
\definecolor{mygrey}{RGB}{105,105,105}
\definecolor{OliveGreen}{RGB}{85,107,47}
\definecolor{NavyBlue}{RGB}{0,0,128}
\definecolor{mygreen}{RGB}{34,139,34}
\definecolor{myY}{RGB}{220,255,203}
\definecolor{myYO}{RGB}{255, 220, 151}

\definecolor{mygreenc}{RGB}{150,50,50}

\usetikzlibrary{patterns.meta}

\newcommand{\Vgate}[2]{
\draw[thick] (#1-0.5, #2 +0.5) -- (#1+0.5,#2-0.5);
\draw[thick] (#1-0.5,#2-0.5) -- (#1+0.5,#2+0.5);
\draw[ thick, fill=myorange, rounded corners=2pt] (#1-0.25,#2+0.25) rectangle (#1+0.25,#2-0.25);
\draw[thick] (#1,#2+0.15) -- (#1+0.15,#2+0.15) -- (#1+0.15,#2);
}

\newcommand{\Vdaggate}[2]{
\draw[thick] (#1-0.5, #2 +0.5) -- (#1+0.5,#2-0.5);
\draw[thick] (#1-0.5,#2-0.5) -- (#1+0.5,#2+0.5);
\draw[ thick, fill=myorange, rounded corners=2pt] (#1-0.25,#2+0.25) rectangle (#1+0.25,#2-0.25);
\pattern[pattern=north west lines, pattern color=black] (#1-0.25,#2+0.25) rectangle (#1+0.25,#2-0.25);
\draw[thick] (#1,#2+0.15) -- (#1+0.15,#2+0.15) -- (#1+0.15,#2);
}

\newcommand{\Vfoldedgate}[2]{
\draw[very thick] (#1-0.5, #2 +0.5) -- (#1+0.5,#2-0.5);
\draw[very thick] (#1-0.5,#2-0.5) -- (#1+0.5,#2+0.5);
\draw[ thick, fill=myred, rounded corners=2pt] (#1-0.25,#2+0.25) rectangle (#1+0.25,#2-0.25);
\draw[thick] (#1,#2+0.15) -- (#1+0.15,#2+0.15) -- (#1+0.15,#2);
}
\newcommand{\Vfoldeddaggate}[2]{
\draw[very thick] (#1-0.5, #2 +0.5) -- (#1+0.5,#2-0.5);
\draw[very thick] (#1-0.5,#2-0.5) -- (#1+0.5,#2+0.5);
\draw[ thick, fill=myred, rounded corners=2pt] (#1-0.25,#2+0.25) rectangle (#1+0.25,#2-0.25);
\pattern[pattern=north west lines, pattern color=black] (#1-0.25,#2+0.25) rectangle (#1+0.25,#2-0.25);
\draw[thick] (#1,#2+0.15) -- (#1+0.15,#2+0.15) -- (#1+0.15,#2);
}

\newcommand{\Vfoldedstargate}[2]{
\draw[very thick] (#1-0.5, #2 +0.5) -- (#1+0.5,#2-0.5);
\draw[very thick] (#1-0.5,#2-0.5) -- (#1+0.5,#2+0.5);
\draw[ thick, fill=myblue, rounded corners=2pt] (#1-0.25,#2+0.25) rectangle (#1+0.25,#2-0.25);
\draw[thick] (#1,#2+0.15) -- (#1+0.15,#2+0.15) -- (#1+0.15,#2);
}

\newcommand{\Vfoldedstardaggate}[2]{
\draw[very thick] (#1-0.5, #2 +0.5) -- (#1+0.5,#2-0.5);
\draw[very thick] (#1-0.5,#2-0.5) -- (#1+0.5,#2+0.5);
\draw[ thick, fill=myblue, rounded corners=2pt] (#1-0.25,#2+0.25) rectangle (#1+0.25,#2-0.25);
\pattern[pattern=north west lines, pattern color=black] (#1-0.25,#2+0.25) rectangle (#1+0.25,#2-0.25);
\draw[thick] (#1,#2+0.15) -- (#1+0.15,#2+0.15) -- (#1+0.15,#2);
}

\usepackage{hyperref}
\hypersetup{
 pdfauthor={Faidon Andreadakis, and Paolo Zanardi},
 pdftitle={Operator Space Entangling Power of Quantum Dynamics and Local Operator Entanglement Growth in Dual-Unitary Circuits},
 pdflang={English},colorlinks=true,linkcolor=RubineRed,citecolor=blue-violet,urlcolor=Cerulean} 
\usepackage[capitalise]{cleveref}

\begin{document}

\title{Operator Space Entangling Power of Quantum Dynamics and Local Operator Entanglement Growth in Dual-Unitary Circuits}

\author{Faidon Andreadakis}
\email [e-mail: ]{fandread@usc.edu}
\affiliation{Department of Physics and Astronomy, and Center for Quantum Information Science and Technology, University of Southern California, Los Angeles, California 90089-0484, USA}

\author{Emanuel Dallas}
\email [e-mail: ]{dallas@usc.edu}
\affiliation{Department of Physics and Astronomy, and Center for Quantum Information Science and Technology, University of Southern California, Los Angeles, California 90089-0484, USA}

\author{Paolo Zanardi}
\email [e-mail: ]{zanardi@usc.edu}
\affiliation{Department of Physics and Astronomy, and Center for Quantum Information Science and Technology, University of Southern California, Los Angeles, California 90089-0484, USA}
\affiliation{Department of Mathematics, University of Southern California, Los Angeles, California 90089-2532, USA}

\date{\today}

\begin{abstract}

Operator entanglement is a well-established measure of operator complexity across a system bipartition. In this work, we introduce a measure for the ability of a unitary channel to generate operator entanglement, representing an operator-level generalization of the state-space entangling power
. This operator space entangling power is demonstrated to be linked to the scrambling properties of the unitary channel via the recently introduced concept of mutual averaged non-commutativity of quantum operator algebras. An upper bound for the operator space entangling power is identified, corresponding to unitary channels with scrambling properties akin to those of typical unitaries
.  Additionally, for Hamiltonian dynamics, we find that the short-time growth rate of the operator space entangling power matches the Gaussian scrambling rate of the bipartite out-of-time-order-correlator, establishing a direct link between information scrambling and operator entanglement generation for short time scales. Finally, we examine the average growth of local operator entanglement across a symmetric bipartition of a spin-chain. For dual-unitary circuits, a combination of analytical and numerical investigations demonstrates that the average growth of local operator entanglement exhibits two distinct regimes in relation to the operator space entangling power of the building-block gate
.
\end{abstract}
\maketitle

\section{Introduction} \label{secintro}

Quantum entanglement is one of the hallmarks of quantum physics that describes non-classical correlations between quantum subsystems and has a central role as a resource in quantum information processing \cite{horodeckiQuantumEntanglement2009}.
At the same time, the scaling of quantum entanglement in quantum many-body systems has been recognized to play a crucial role in the efficiency of certain representations of quantum states. Ground states of locally interacting gapped Hamiltonians exhibit low levels of entanglement,  area law entanglement \cite{eisertColloquiumAreaLaws2010}, which exactly underpins the success of tensor network methods in efficiently describing them \cite{verstraeteMatrixProductStates2008,orusPracticalIntroductionTensor2014,feldmanEntanglementEstimationTensor2022}.
\par The concept of entanglement can be extended in operator space utilizing the Hilbert space structure of the space of linear operators. This notion is the \emph{operator entanglement} originally defined in Refs. \cite{wangQuantumEntanglementUnitary2002,zanardiEntanglementQuantumEvolutions2001}
with regard to the unitary evolution operator and is related to the complexity of an operator across a bipartition. Complex quantum dynamics generate nonlocal correlations, which, in the Heisenberg picture, correspond to the growth of the support of initially local operators. This leads to the spreading of information across the system degrees of freedom, a process termed information scrambling \cite{xuScramblingDynamicsOutofTimeOrdered2024,shenkerBlackHolesButterfly2014,haydenBlackHolesMirrors2007,
robertsChaosComplexityDesign2017,mezeiEntanglementSpreadingChaotic2017,lashkariFastScramblingConjecture2013}. Remarkably, the operator entanglement of the unitary evolution operator is directly linked with information scrambling and entropy production and its system size scaling has been observed to contain signatures of integrability breaking \cite{styliarisInformationScramblingBipartitions2021}. The concept of operator entanglement has also proven useful in characterizing the classical simulability of quantum evolutions by considering the dynamics of the operator entanglement of initially local observables under Heisenberg evolution \cite{prosenEfficiencyClassicalSimulations2007}, or \emph{local operator entanglement}. This growth of local operator entanglement is a probe of the dynamical complexity of quantum many-body systems \cite{prosenOperatorSpaceEntanglement2007,pizornOperatorSpaceEntanglement2009,
znidaricComplexityThermalStates2008,
muthDynamicalSimulationIntegrable2011,dubailEntanglementScalingOperators2017}, with the latter being considered a key signature of quantum chaos \cite{prosenChaosComplexityQuantum2007,jonayCoarsegrainedDynamicsOperator2018}. We should note that quantum chaotic models necessarily exhibit information scrambling \cite{dowlingScramblingNecessaryNot2023c},
but the converse is not always true \cite{xuDoesScramblingEqual2020,dowlingScramblingNecessaryNot2023c}.
\par Given a bipartite system, a natural question arises about the ability of quantum dynamics to generate quantum entanglement. This motivates the definition of \emph{entangling power}, namely the on-average capacity of a unitary channel to entangle initially product quantum states \cite{zanardiEntanglingPowerQuantum2000}. This is a distinct, albeit related,
quantity to operator entanglement \cite{zanardiEntanglementQuantumEvolutions2001} and has been used in the study of integrability and chaos in quantum many-body systems \cite{wangEntanglementSignatureQuantum2004,palEntanglingPowerTimeevolution2018,haakeQuantumSignaturesChaos2010}. Recently, it was shown that the entangling power correlates with the ergodic properties of dual-unitary quantum circuit dynamics \cite{aravindaDualunitaryQuantumBernoulli2021}. Apart from their significance in quantum computing, local quantum circuits provide a model that simulates Trotterized Hamiltonian evolutions \cite{lloydUniversalQuantumSimulators1996a,ortizQuantumAlgorithmsFermionic2001}. The dual-unitarity condition imposes the dynamics to be unitary both in space and time and allows the explicit calculation of spectral and correlation functions \cite{bertiniExactCorrelationFunctions2019,bertiniExactSpectralForm2018,kosCorrelationsPerturbedDualUnitary2021} and entanglement spreading \cite{bertiniEntanglementSpreadingMinimal2019}, while maintaining a rich ergodic structure \cite{bertiniExactCorrelationFunctions2019}.
\par In this paper, we extend the concept of entangling power to the operator level. To this end, we study the average generation of operator entanglement induced by a unitary channel, which we refer to as \emph{operator space entangling power}.
This quantity is a characteristic of the unitary evolution itself and is independent of any specific choice of operators. In \cref{sec:preliminaries}, we review the concepts of operator entanglement \cite{zanardiEntanglementQuantumEvolutions2001} and mutual averaged non-commutativity of operator algebras \cite{zanardiMutualAveragedNoncommutativity2024a}. In \cref{sec:opentp}, we introduce the operator space entangling power, which we study analytically, demonstrating its connection to the mutual averaged non-commutativity of the subsystem subalgebras and providing an upper bound. In \cref{sec:app}, we investigate operator space entangling power for dynamics given by quantum many-body Hamiltonians and dual-unitary local quantum circuits. Finally, \cref{sec:conclusion} provides conclusions and avenues for future research. Derivations of the reported results and details about the numerical simulations are organized in the Appendix.
%
%


\section{Preliminaries} \label{sec:preliminaries}
Consider a quantum system represented by a $d$-dimensional Hilbert space $\mathcal{H}$. The space of linear operators $\mathcal{L}(\mathcal{H})$ endowed with the Hilbert-Schmidt inner product $\langle X , Y \rangle \coloneqq \Tr( X^\dagger Y)$ is itself a Hilbert space $\mathcal{H}_{HS}$ that is isomorphic to $\mathcal{H}^{\otimes 2}$. This allows us to assign a state to every unitary operator $U\in \mathcal{L}(\mathcal{H})$ as $\ket{U}=U\otimes \mathds{1} \ket{\Phi^+}$, where $\ket{\Phi^+} \coloneqq 1/\sqrt{d} \sum_{i=1}^d \ket{i} \otimes \ket{i}$ with $\{\ket{i}\}_{i=1}^d$ an orthonormal basis of $\mathcal{H}$. Assuming a bipartite structure on $\mathcal{H}\cong \mathcal{H}_{A} \otimes \mathcal{H}_B$, we have for every $U$ the associated state $\ket{U}_{ABA^\prime B^\prime} = 1/\sqrt{d} \sum_{i=1}^{d_A} \sum_{j=1}^{d_B} U \ket{ij} \otimes \ket{ij}$. The entanglement of $U$ is then defined to be the entanglement of this state across the $AA^\prime \vert BB^\prime$ bipartition. Using the linear entropy of the reduced state $\sigma_{AA^\prime} \coloneqq \Tr_{BB^\prime}(\ket{U}_{ABA^\prime B^\prime}\bra{U})$ as the entanglement measure, one gets the definition of \emph{operator entanglement} \cite{zanardiEntanglementQuantumEvolutions2001}: $E(U)\coloneqq 1- \Tr_{AA^\prime}( \sigma_{AA^\prime}^2)$. 
\par In Ref. \cite{zanardiEntanglementQuantumEvolutions2001} it was shown that
\begin{equation} \label{opent}
E(U)=1-\frac{1}{d^2} \Tr(\mathsf{S}_{AA^\prime} U^{\otimes 2} \mathsf{S}_{AA^\prime} {U^\dagger}^{\otimes 2}),
\end{equation}
where $\mathsf{S}_{AA^\prime}$ denotes the swap between the $A$ and $A^\prime$ subsystems and \cref{opent} remains identical if we substitute $A$ with $B$. Remarkably, the operator entanglement is exactly equal to a type of out-of-time-order correlator (OTOC) for dynamics given by the unitary $U$ \cite{styliarisInformationScramblingBipartitions2021},
\begin{equation} \label{bipartite-otoc}
\begin{split}
&E(U)=\frac{1}{2d}\mathlarger{\mathbb{E}}_{X_A,Y_B} \left[ \left\Vert \left[ \mathcal{U}(X_{A}), Y_B \right] \right\Vert_2^2 \right]=\\
&=1-\frac{1}{d} \Re{\mathlarger{\mathbb{E}}_{X_A,Y_B} \left[ \Tr(\mathcal{U}(X_{A}^\dagger)\,  Y_B^\dagger \, \mathcal{U}(X_{A}) \, Y_B) \right]}.
\end{split}
\end{equation}
Here $\mathcal{U}(X_A) \equiv U \, X_A \, U^\dagger$ and $\mathbb{E}_{X_A,Y_B}$ denotes averaging over the Haar measure on the unitary subgroups of operators acting on each of the subsystems. The second line follows by expanding the 2-norm $\Vert M \Vert_2^2 \coloneqq \Tr(X^\dagger X)$ and using the fact that $X_A$, $Y_B$ are chosen to be unitaries. This quantity describes the on-average non-commutativity of Heisenberg evolved operators initially localized in subsystem $A$ with operators in the subsystem $B$ and probes the scrambling of information induced by $U$ between the two subsystems.
\par The above measure of non-commutativity can be utilized in a more general algebraic setting \cite{zanardiQuantumScramblingObservable2022,andreadakisScramblingAlgebrasOpen2023,
zanardiMutualAveragedNoncommutativity2024a}. Specifically, given two hermitian-closed unital operator algebras $\mathcal{A}$, $\mathcal{B}$ one can define their mutual averaged non-commutativity as \cite{zanardiMutualAveragedNoncommutativity2024a}
\begin{equation} \label{man}
S(\mathcal{A}:\mathcal{B})\coloneqq \frac{1}{2d}\mathlarger{\mathbb{E}}_{X \in \mathcal{A},Y \in \mathcal{B}} \left[ \left\Vert \left[ X, Y \right] \right\Vert_2^2 \right],
\end{equation}
where, similar to before, $X$ and $Y$ are drawn from the Haar distribution on the unitary subgroups of operators in $\mathcal{A}$ and $\mathcal{B}$, respectively. Clearly, identifying as $\mathcal{A}$, $\mathcal{B}$ the subalgebras of operators acting on the subsystems $A$, $B$, \cref{bipartite-otoc} is equivalent to
\begin{equation} \label{man_opent}
E(U)=S(\mathcal{U}(\mathcal{A}):\mathcal{B}).
\end{equation}
Intuitively, this is a simple measure that quantifies to what degree two algebras fail to commute, and in the quantum information setting, this non-commutativity is exactly the origin of the quantum scrambling induced by $\mathcal{U}$.

\section{Operator Space Entangling Power} \label{sec:opentp}
Let us introduce the main object of this work. Given a bipartite Hilbert space $\mathcal{H} \cong \mathcal{H}_A \otimes \mathcal{H}_B$ and a unitary channel $\mathcal{U}[\bullet ] = U [\bullet ] U^\dagger$, we define the \emph{operator space entangling power}
as
\begin{equation}\label{opentp_def}
E_p(\mathcal{U})=\mathlarger{\mathbb{E}}_{X_A,Y_B} E\left(\mathcal{U}(X_A \otimes Y_B)\right),
\end{equation}
where $X_A$, $Y_B$ are Haar random unitary operators on the subsystems $A$, $B$. This quantity describes how much operator entanglement is generated by $\mathcal{U}$ when acting on initially unentangled product operators $X_A \otimes Y_B$.
\par Similarly as before, denote the subsystem algebras as $\mathcal{A} \cong \mathcal{L}(\mathcal{H}_A)\otimes \mathds{1}_B$, $\mathcal{B} \cong \mathds{1}_A \otimes \mathcal{L}(\mathcal{H}_B)$. Substituting \cref{opent} in \cref{opentp_def} and performing the average analytically, see \cref{app:proofs_1}, we get
\begin{widetext}
\begin{equation} \label{opentp_general}
E_p(\mathcal{U})= N(d_A,d_B)\left[1-\left(1-\frac{S(\mathcal{U}(\mathcal{A}):\mathcal{B})}{S(\mathcal{A}:\mathcal{A})}\right)^2-\frac{d_A^2}{d_B^2} \left( 1-\frac{S(\mathcal{U}(\mathcal{A}):\mathcal{A})}{S(\mathcal{A}:\mathcal{A})}\right)^2-\frac{2}{d_B^2}\frac{S(\mathcal{U}(\mathcal{A}):\mathcal{B})}{S(\mathcal{A}:\mathcal{A})}\frac{S(\mathcal{U}(\mathcal{A}):\mathcal{A})}{S(\mathcal{A}:\mathcal{A})}\right],
\end{equation}
\end{widetext}
where $N(d_A,d_B)\coloneqq \frac{d_B^2}{d_A^2} \frac{d_A^2-1}{d_B^2-1}$ and for $X,Y \in \{A,B\}$ the mutual-averaged non-commutativities are given as \cite{zanardiMutualAveragedNoncommutativity2024a}
\begin{equation}
S(\mathcal{U}(\mathcal{X}):\mathcal{Y}) \equiv 1-\frac{1}{d}\Tr(\mathsf{S} \, \mathcal{U}\left(\frac{\mathsf{S}_{XX^\prime}}{d_X}\right)\frac{\mathsf{S}_{YY^\prime}}{d_Y}).
\end{equation}
In the above equation, $\mathsf{S}$ is the swap between the two copies $AB$ and $A^\prime B^\prime$ of the Hilbert space and $\mathsf{S}_{AA^\prime}, \mathsf{S}_{BB^\prime}$ are the swaps between the subsystem copies. The same expression in \cref{opentp_general} holds if we exchange the roles of $A$ and $B$ and without loss of generality we can assume that $d_B \geq d_A$. Then, the ratios $\frac{S(\mathcal{U}(\mathcal{A}):\mathcal{B})}{S(\mathcal{A}:\mathcal{A})},\frac{S(\mathcal{U}(\mathcal{A}):\mathcal{A})}{S(\mathcal{A}:\mathcal{A})} \in [0,1]$ can be thought of as the probabilities that an evolved element of $\mathcal{A}$ does not commute with some element of $\mathcal{B}$ and $\mathcal{A}$, respectively \cite{zanardiMutualAveragedNoncommutativity2024a}. As we noted in \cref{man_opent}, the mutual averaged non-commutativity $S(\mathcal{U}(\mathcal{A}):\mathcal{B})$ equals to the operator entanglement $E(U)$ which is related to the scrambling of information across the $A:B$ bipartition. In passing, let us note that \cref{opentp_def} may be interpreted as the effect of local unitaries on the operator entanglement of a composite unitary evolution $U (U_A \otimes U_B) V$ for $V=U^\dagger$. The role of local unitaries in these types of evolutions has been investigated in Ref. \cite{jonnadulaEntanglementMeasuresBipartite2020} for quantities related to the operator entanglement, obtaining expressions in agreement to \cref{opentp_general}.
\par In Ref. \cite{styliarisInformationScramblingBipartitions2021}, $E(U)$ was expressed in terms of an average subsystem entropy production as
\begin{align}
S(\mathcal{U}(\mathcal{A}):\mathcal{B})=\frac{d_A+1}{d_A} \mathlarger{\mathbb{E}}_{\ket{\psi}} S_{lin}\left[ \Tr_B(\mathcal{U}(\rho_\psi )) \right], \label{entropy_1}
\end{align}
where $\rho_\psi \coloneqq \ketbra{\psi}{\psi} \otimes \mathds{1}_B/d_B$, $S_{lin}[\rho ] \coloneqq 1 - \Tr(\rho^2)$ is the linear entropy and $\mathlarger{\mathbb{E}}_{\ket{\psi}}$ denotes the average over Haar random states $\ket{\psi}$. We can extend this intuition to $S(\mathcal{U}(\mathcal{A}):\mathcal{A})$, see \cref{app:proofs_1},
\begin{align}
&S(\mathcal{U}(\mathcal{A}):\mathcal{A})=\frac{d_A+1}{d_A} \left( \frac{d_B}{d_A} \mathlarger{\mathbb{E}}_{\ket{\psi}} S_{lin}\left[ \Tr_A( \mathcal{U}(\rho_\psi)) \right] + \right.\nonumber\\
&\hspace{120pt}\left. +1-\frac{d_B}{d_A} \right). \label{entropy_2}
\end{align}
By construction, the quantum state $\rho_\psi$ is pure in the subsystem $A$ (zero entropy) and maximally mixed in the subsystem $B$ (maximal entropy). \cref{entropy_1,entropy_2} describe, respectively, the on-average entropy generated in subsystem $A$ and entropy remaining in subsystem $B$ after the action of the unitary channel $\mathcal{U}$ on $\rho_\psi$. 
In general, the entropy increase in subsystem $A$ and entropy decrease in subsystem $B$ need not be equal due to the generation of quantum entanglement under $\mathcal{U}$.

\par In the important special case of a symmetric bipartition, where $d_A = d_B =\sqrt{d}$, we see that $S(\mathcal{U}(\mathcal{A}):\mathcal{A})=E(US)$. The operator space entangling power (\cref{opentp_general}) then reduces to a simple function of the operator entanglement ratios $E(U)/E(\mathsf{S})$ and $E(U\mathsf{S})/E(\mathsf{S})$:
\begin{equation} \label{opentp_sym}
\begin{split}
E_p(\mathcal{U})=&1-\left(1-\frac{E(U)}{E(\mathsf{S})}\right)^2-\left( 1-\frac{E(U\mathsf{S})}{E(\mathsf{S})} \right)^2 - \\
&- \frac{2}{d} \frac{E(U)}{E(\mathsf{S})} \frac{E(U\mathsf{S})}{E(\mathsf{S})}.
\end{split}
\end{equation}
Also, \cref{entropy_1,entropy_2} reduce to
\begin{equation} \label{entropy_reduce}
\begin{split}
&E(U)=\frac{\sqrt{d}+1}{\sqrt{d}} \mathlarger{\mathbb{E}}_{\ket{\psi}} S_{lin}\left[ \Tr_B(\mathcal{U}(\rho_\psi )) \right],\\
&E(U\mathsf{S})=\frac{\sqrt{d}+1}{\sqrt{d}} \mathlarger{\mathbb{E}}_{\ket{\psi}} S_{lin}\left[ \Tr_A(\mathcal{U}(\rho_\psi )) \right],
\end{split}
\end{equation}
which shows that, up to a constant prefactor, $E(U)$ and $E(U\mathsf{S})$ are equal to the average linear entropy  of the evolved states $\mathcal{U}(\rho_\psi )$ in the subsystems $A$ and $B$, respectively.

\subsection{Hamiltonian Dynamics and Short-time Growth} \label{subsec:short}
Let $U_t=\exp(itH)$ be the one-parameter family of unitaries associated with the dynamics generated by the Hamiltonian $H$. In Ref. \cite{zanardiOperationalQuantumMereology2024} it was shown that
\begin{equation} \label{short_bip}
\begin{split}
&E(U_t)=S(\mathcal{U}_t(\mathcal{A}):\mathcal{B})=2\left(\frac{t}{\tau_s} \right)^2 + O(t^3),\\
&\tau_s^{-1}= \frac{1}{\sqrt{d}}\left\lVert{H-\frac{\mathds{1}_A}{d_A} \otimes \Tr_A(H) -\Tr_B(H)\otimes\frac{\mathds{1}_B}{d_B}}\right\rVert_2.
\end{split}
\end{equation}
$\tau_s^{-1}$ is the Gaussian scrambling rate that dictates the short-time behavior of the operator entanglement of $U_t$. It turns out that this rate also dictates the short-time behavior of the operator space entangling power. This follows from the observation that $S(\mathcal{U}_t(\mathcal{A}):\mathcal{A})$ is independent of the Hamiltonian up to order $O(t^2)$. Specifically, see \cref{app:short},
\begin{equation} \label{observation}
 S(\mathcal{U}_t(\mathcal{A}):\mathcal{A})= 1 -\frac{1}{d_A^2}+ O(t^3).
 \end{equation}
Combining \cref{observation} with \cref{opentp_general,short_bip}, we get
\begin{equation} \label{opentp_short}
E_p(\mathcal{U}_t) = 4 \left(\frac{t}{\tau_s} \right)^2 + O(t^3).
\end{equation}
In Ref. \cite{zanardiOperationalQuantumMereology2024} the connection between $E(U_t)$ and scrambling was utilized to define ``informationally stable'' subsystems with respect to given Hamiltonian dynamics as those that minimize the Gaussian rate $\tau_s^{-1}$. This corresponds to a criterion for quantum mereology \cite{carrollQuantumMereologyFactorizing2021}, namely the partitioning of a quantum system into subsystems. \cref{opentp_short} shows that these same subsystems also minimize the generation of operator entanglement by $\mathcal{U}_t$ for short timescales. In fact, for $t\ll\tau_s$ we have $E_p(\mathcal{U}_t) \sim 2 E(U_t)$; in the early growth regime, the operator space entangling power of $\mathcal{U}_t$ is exactly proportional to the operator entanglement of $U_t$.
\subsection{Upper bound, typical unitaries and entangling power} \label{subsec:bound}

To gain more intuition about the properties of the operator space entangling power, it is useful to consider its extremal points. Naturally, the lowest value can always be achieved by the trivial evolution $U=\mathds{1}$, whence $E_p (\mathcal{I})=0$. Also, in the case of a symmetric bipartition, for $U=\mathsf{S}$ we have $E_p(\mathcal{S})=0$, despite $E(\mathsf{S})$ being maximal. This is a simple case where entanglement generation is distinguished from scrambling. The swap dynamics are sufficient to transport information between the two subsystems by exchanging the states between the subsystems $A$ and $B$. In terms of \cref{entropy_reduce}, this type of process maximally increases the entropy in $A$, while reducing the entropy in $B$ to zero. On the contrary, entanglement generation requires a ``non-local'' dispersion of information such that both average entropies in \cref{entropy_reduce} are non-zero. Treating $S(\mathcal{U}(\mathcal{A}):\mathcal{B})$, $S(\mathcal{U}(\mathcal{A}):\mathcal{A})$ as independent variables, we can obtain a simple upper bound for $E_p(\mathcal{U})$, see \cref{app:max},
\begin{equation} \label{opentp_bound}
E_p(\mathcal{U}) \leq \frac{(d_A^2-1)(d_B^2-1)}{d^2-1}.
\end{equation}
Note that $\mathbb{E}_U E(U) = (d_A^2-1)(d_B^2-1)/(d^2-1)$, where the average is done with respect to Haar random unitaries $U$ \footnote{On account of \cref{opent} it is sufficient the distribution of $U$ to be a unitary 2-design \cite{robertsChaosComplexityDesign2017}.} \cite{styliarisInformationScramblingBipartitions2021}. Intuitively, the upper bound in \cref{opentp_bound} is approached by unitary channels that map the Haar random product operators $X_A \otimes Y_B$ to sufficiently uniformly distributed operators in the full space, such that the average in \cref{opentp_def} approaches the Haar average value. More precisely, the bound is achieved if there exists a unitary $U_{\star}$ such that 
\begin{equation} \label{max_condition}
\begin{split}
&S(\mathcal{U}_{\star}(\mathcal{A}):\mathcal{B})= \mathbb{E}_U S(\mathcal{U}(\mathcal{A}):\mathcal{B}) = \frac{(d_A^2-1)(d_B^2-1)}{d^2-1},\\ 
&S(\mathcal{U}_{\star}(\mathcal{A}):\mathcal{A})=\mathbb{E}_U S(\mathcal{U}(\mathcal{A}):\mathcal{A}) = \frac{d_B^2}{d_A^2} \frac{(d_A^2-1)^2}{d^2-1}.
\end{split}
\end{equation}
This means that $U_{\star}$ should approximate the properties of a typical ensemble in terms of the mutual averaged non-commutativities $S(\mathcal{U}(\mathcal{A}):\mathcal{B})$, $S(\mathcal{U}(\mathcal{A}):\mathcal{A})$.
\par Let us focus on the case of a symmetric bipartition, $d_A=d_B=\sqrt{d}$. Then, \cref{opentp_bound} reduces to 
\begin{equation} \label{reduce_bound}
E_p(\mathcal{U}) \leq 1-\frac{2}{d+1}
\end{equation}
and the maximality condition is $E(U_{\star})=E(U_{\star} S)=1-2/(d+1)$. In general, it is non-trivial to check whether this condition can be satisfied, since $E(U)$ and $E(US)$ are not independent variables. For the case of 2 qubits $d_A=d_B=2$, their feasible domain has been investigated in Ref. \cite{youEntanglementFeaturesRandom2018}. Defining $I_1 \coloneqq 1-4/3 \, E(U)$, $I_2 \coloneqq 1- 4/3\, E(US)$, it was found that
\begin{equation} \label{domain}
\begin{split}
&I_1 + I_2 \geq 1/3 \\
&\sqrt{I_1} + \sqrt{I_2} \leq 1.
\end{split}
\end{equation}
The above bounds are tight in the sense that there exist 2-qubit unitaries along the boundaries. The condition for $U_{\star}$ corresponds to $I_1 = I_2 =1/5$, which is well within the feasible domain \cite{akhtarDualUnitaryClassicalShadow2024}.
\par For the case of $d_A=d_B>2$ the exact constraints on the feasible domain are not known. Interestingly, though, for this case there exists a class of unitaries, called perfect tensors \cite{pastawskiHolographicQuantumErrorcorrecting2015} or 2-unitaries \cite{goyenecheAbsolutelyMaximallyEntangled2015,ratherConstructionLocalEquivalence2022a}, for which both $E(U)$ and $E(US)$ are maximum and equal to $E(S)=1-1/d$. Perfect tensors do not exist for the qubit case and it can be readily confirmed that they would violate the constraints in \cref{domain}. The operator space entangling power of the perfect tensors is $E_p(U)=1-2/d$, which is not maximal, but approaches the upper bound of \cref{reduce_bound} in large dimensions. In order to further evaluate the tightness of this upper bound for small dimensions, we numerically search for unitaries that maximize the operator space entangling power using a gradient ascent algorithm, see \cref{appendix_gradient}. Running the algorithm \cite{Andreadakis_Python_Code_Gradient} for $2\leq \sqrt{d} \leq 8$, we observe that we can always achieve the upper-bound in \cref{reduce_bound} within machine precision error.
\par We should note that the operator space entangling power introduced in this paper corresponds to an operator level generalization of a related quantity that probes the ability of a unitary channel to generate \emph{state-space} entanglement. This \emph{entangling power} is defined as 
\begin{equation}
e_p(\mathcal{U}) \coloneqq \mathbb{E}_{\ket{\psi},\ket{\phi}} S_{lin} \left[\Tr_A(\mathcal{U}(\ketbra{\psi}{\psi} \otimes \ketbra{\phi}{\phi}))\right],
\end{equation}
where $\ket{\psi}$, $\ket{\phi}$ are Haar random pure states on $\mathcal{H}_A$, $\mathcal{H}_B$ \cite{zanardiEntanglingPowerQuantum2000}. For a symmetric bipartition $d_A=d_B$, $e_p(\mathcal{U})$ is also a function of $E(U)$, $E(US)$,
\begin{equation}
e_p(\mathcal{U}) = \frac{d}{(\sqrt{d}+1)^2}(E(U)+E(US)-E(S)).
\end{equation}
Similarly to what we noted for the operator space entangling power, for $U=S$ we have $e_p(\mathcal{S})=0$, despite $E(S)$ being maximal, since the swap operator transports information from $A$ to $B$ in a manner that does not generate entanglement. In fact, on account of \cref{entropy_reduce}, $e_p(\mathcal{U})$ is exactly proportional to the difference between the average entropy increase in subsystem $A$ and entropy decrease in subsystem $B$ when $\mathcal{U}$ acts on the states $\rho_\psi = \ketbra{\psi}{\psi}\otimes \mathds{1}_B/\sqrt{d}$. In the case of 2 qubits, where the constraints of \cref{domain} apply, $e_p(\mathcal{U})$ is maximized for $I_1=1/3,I_2=0$, e.g., the $\pi/4-ZZ$ gate $e^{i\pi/4\sigma_z \otimes \sigma_z}$, or $I_1=0,I_2=1/3$, e.g., the iSWAP gate $e^{i\pi/4(\sigma_x \otimes \sigma_x + \sigma_z \otimes \sigma_z)}$ \cite{akhtarDualUnitaryClassicalShadow2024}. For $d_A=d_B>2$, $e_p(\mathcal{U})$ is maximal for the class of perfect tensors mentioned above. These maximality conditions are distinct from those of $E_p(\mathcal{U})$. Intuitively, $e_p(\mathcal{U})$ is maximal when the subsystem entropies, in terms of \cref{entropy_reduce}, are maximized, whereas $E_p(\mathcal{U})$ is maximal when product operators $X_A \otimes Y_B$ are mapped to sufficiently random operators.

\section{Quantum Many-body Dynamics \& Local Quantum Circuits} \label{sec:app}

\begin{figure*}
\centering
\begin{subfigure}{.5\textwidth}
  \centering
  \includegraphics[width=1\linewidth]{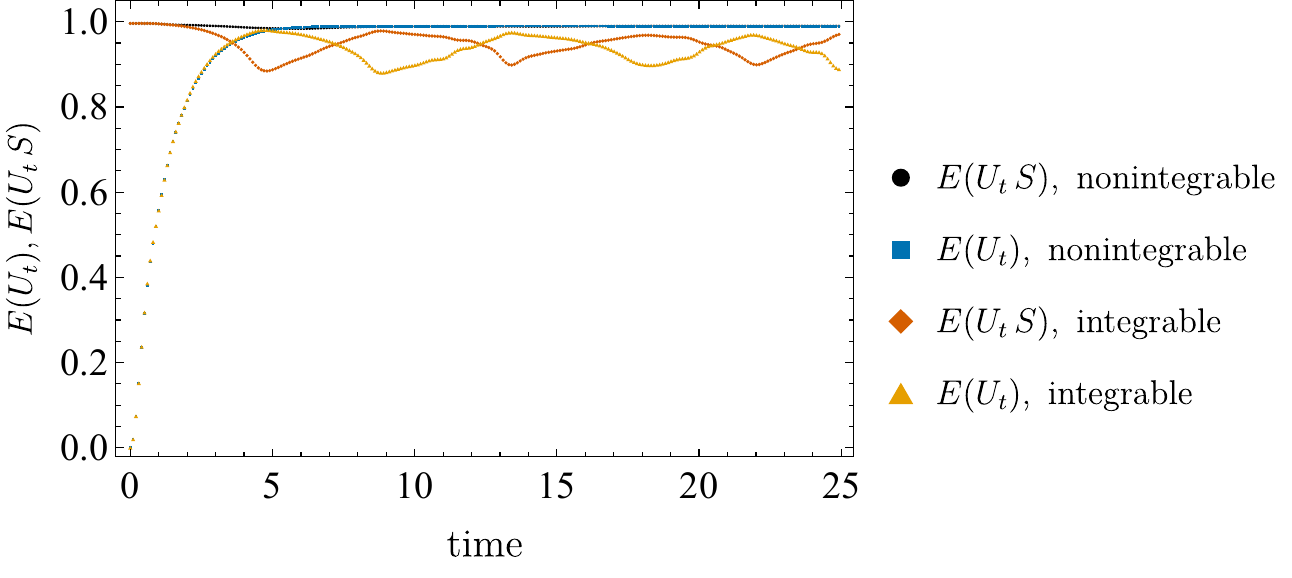}
 \caption{}\label{fig:nonint_int}
\end{subfigure}%
\begin{subfigure}{.5\textwidth}
  \centering
  \includegraphics[width=1\linewidth]{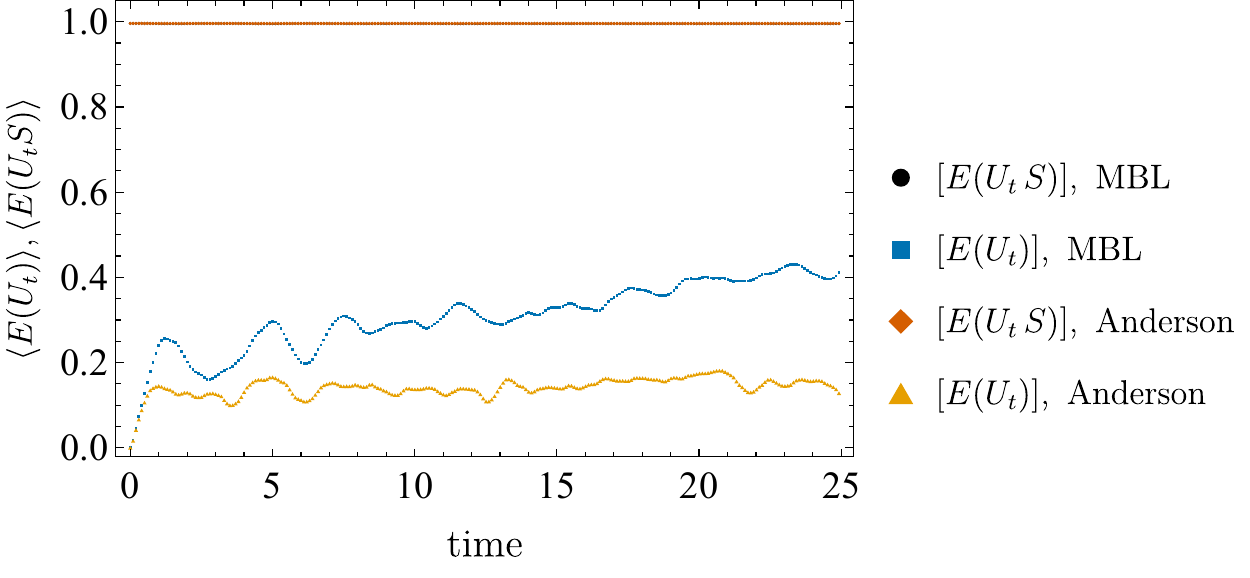}
  \caption{}\label{fig:localized}
\end{subfigure}
\caption{The time dynamics of $E(U_t)$ and $E(U_tS)$ for the Hamiltonian models discussed in the main text. (a) For the nonintegrable TFIM model both $E(U_t)$ and $E(U_tS)$ quickly equilibrate to a near maximal value, while in the integrable case $E(U_t)$ and $E(U_tS)$ oscillate perfectly anti-phased around a slightly lower equilibration value. (b) For the localized models the growth of $E(U_t)$ is suppressed, while $E(U_tS)$ remains near its initial value at all times.}

\label{fig:comp}
\end{figure*}

In this section, we will investigate the role of operator space entangling power in physical applications and its relationship to previously studied quantities and phenomena. In doing so, we will focus on numerical simulations of prototypical models of quantum evolution. Although we will be selecting specific systems and model parameters, we expect that the qualitative nature of our observations will apply to a broader range of situations.

\subsection{Hamiltonian evolution, Integrability \& Localization} \label{subsec:QMB}

Consider a 1D spin-1/2 chain with an even number of sites, $L$, and dynamics given by a Hamiltonian $H$ with open boundary conditions. The Hilbert space is naturally equipped with a tensor product structure $\mathcal{H} \cong \otimes_{i=1}^L \mathcal{H}_i$, where $\mathcal{H}_i \cong \mathbb{C}^2$ is the qubit Hilbert space and $\dim(\mathcal{H}) = d =2^L$. Given a Hamiltonian $H$, we will be discussing the entangling properties of the unitary operator $U_t=\exp(iHt)$ across a half-half bipartition, looking at qualitative features of the dynamics beyond the short-time regime discussed in \cref{subsec:short}.
\par The models we will be examining have the general form
\begin{equation}\label{hamiltonian}
H=-\sum_{i=1}^{L-1} \sigma_i^z \sigma_{i+1}^z - \sum_{i=1}^L \left(h \, \sigma_i^z + g_i \, \sigma_i^x\right),
\end{equation}
where $h,g_i$ are constants and $\sigma_i^{(x,z)}$ denote Pauli operators on site $i$. Different selections for the coupling constants lead to different behavior for the quantum dynamics \cite{anandBROTOCsQuantumInformation2022}. Specifically, we consider the following cases: (i) $h=0.5$, $g_i=1.05 \, \forall \, i$, corresponding to a nonintegrable point of the transverse-field Ising model (TFIM) with on-site magnetization, (ii) $h=0$, $g_i =1 \, \forall \, i$, corresponding to an integrable point of the TFIM model, (iii) $h=0$, $g_i \in [-10,10]$, drawn from the uniform distribution corresponding to an Anderson localized model, (iv) $h=0.5$, $g_i \in [-10,10]$, drawn from the uniform distribution corresponding to a many-body localized (MBL) model. Note that these coupling parameters set the energy scale and, thus, the relevant timescales of the dynamics.

\par We numerically simulate the time evolution operator via exact diagonalization for a system of $L=8$ qubits and compute $E(U_t)$ and $E(U_tS)$ as a function of time for each model. Let us review first the temporal behavior of the functions $E(U_t)$ and $E(U_tS)$ themselves. In \cref{fig:nonint_int} we compare the integrable and nonintegrable point of the TFIM. For the nonintegrable model both $E(U_t)$ and $E(U_tS)$ quickly equilibrate to a value near the maximal one, while for the integrable model $E(U_t)$ and $E(U_tS)$ oscillate around a slightly lower equilibration value. While in both cases $E(U_t)$ grows quickly for short times, the oscillations present in the second case are a standard signature of integrability. What is more, the oscillations of $E(U_t)$ and $E(U_tS)$ are exactly anti-phased, which in light of \cref{entropy_reduce} shows that there is a partial ``exchange'' of entropy between the two halves of the chain. This phenomenon is associated with the symmetries and local charges of motion \cite{gradyInfiniteSetConserved1982} of the integrable TFIM. We note that for more complicated models, for example integrable models by Bethe ansatz that cannot be mapped to a free model, the dynamical signatures of integrability may be more subdued. In \cref{fig:localized} we plot $[ E(U_t)]$ and $[ E(U_tS) ]$ for the localized models, where the square brackets $[ \bullet ]$ denote the disorder average. We observe that in both models the growth of $E(U_t)$ is significantly suppressed, while $E(U_tS)$ is virtually constant and equal to its initial value. On account of \cref{entropy_reduce}, for initial states of the form $\rho_\psi = \ketbra{\psi}{\psi}\otimes \mathds{1}_B/\sqrt{d}$, both models exhibit reduced entropy production in system $A$ and minimal entropy decrease in system $B$. This indicates that the information remains largely localized and any coherent transport of information is absent.
\par
\begin{figure}
\centering
\begin{subfigure}{.5\textwidth}
  \centering
  \includegraphics[width=1\linewidth]{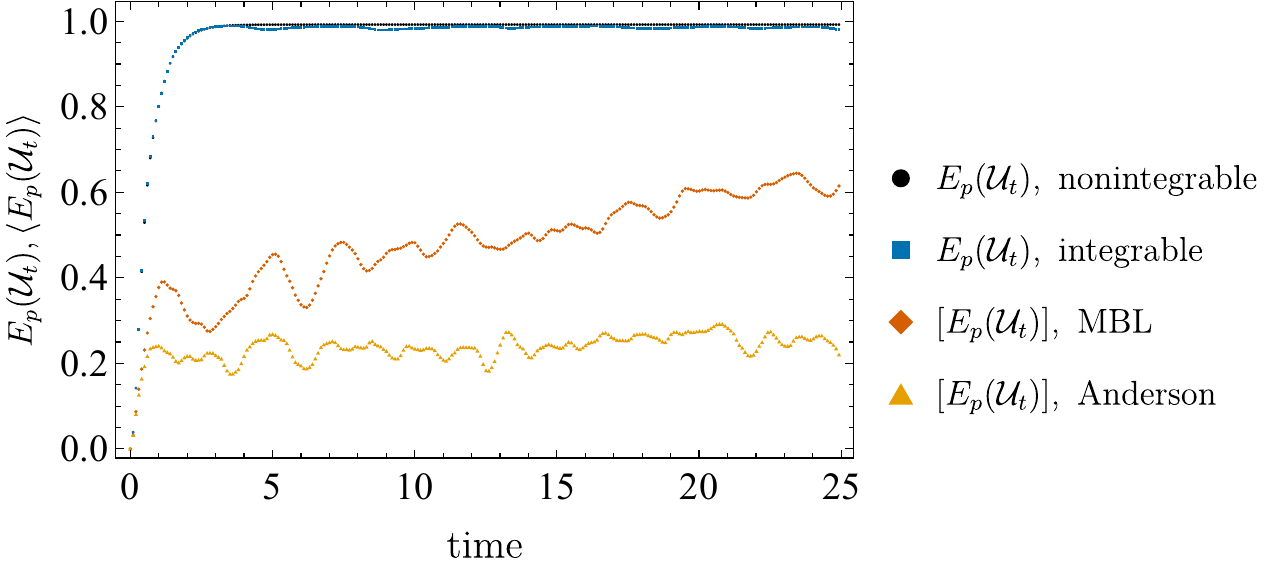}
\end{subfigure}%
\caption{Comparison of the operator space entangling power of the four models as a function of time. The localized models exhibit a clearly distinct behavior, where the generation of operator entanglement is suppressed even at late times. The operator space entangling power of both the integrable and nonintegrable TFIM quickly grows to a large equilibration value, with the behavior of the two curves being largely indistinguishable.}

\label{fig:opentp}
\end{figure}

 Let us now compare the time dynamics of the operator space entangling power for the four models. \cref{fig:opentp} shows that the generation of operator entanglement in the localized phases grows at a slow rate, with the operator space entangling power remaining suppressed even at later times. Nonetheless, any signature of integrability breaking in the TFIM is largely obscured when considering the operator space entangling power. While this may seem counterintuitive, we attribute it to the requirement that the operators $X_A$, $Y_B$ must be drawn from at least a unitary 2-design to reproduce the Haar average in \cref{opentp_def} \cite{robertsChaosComplexityDesign2017}. Such operators are highly non-local within the subsystems $A$ and $B$, rendering the operator space entangling power insensitive to the underlying multi-site tensor product structure. On the contrary the average in \cref{bipartite-otoc} for $E(U)$ may be faithfully reproduced by operators drawn from a unitary 1-design that factorizes completely over the tensor product structure \cite{styliarisInformationScramblingBipartitions2021}. This locality structure is inherently crucial to the notion of integrability, e.g. in terms of local conserved charges. This observation aligns with previous works that have linked integrability and classical simulability to the growth of operator entanglement by examining the operator entanglement dynamics of initially ultra-local operators\footnote{The term ultra-local is used here to emphasize that the support of these operators is fixed irrespective of the system size. This means that the operators are not simply local with respect to subsystems whose size may scale with the system size nor they consist of extensive sums of local terms.}\cite{prosenEfficiencyClassicalSimulations2007,muthDynamicalSimulationIntegrable2011,
dubailEntanglementScalingOperators2017}.\\

\par 

\subsection{Dual Unitary Circuits \& Local Operator Entanglement Growth} \label{subsec:DU}

We will now focus on dynamics generated by local quantum circuits. We consider a system that consists of an even number of sites, $L$, with open boundary conditions and $q$-dimensional local Hilbert spaces. As so, the total Hilbert space is $\mathcal{H} \cong \otimes_{i=1}^L \mathcal{H}_i$ where each $\mathcal{H}_i \cong \mathbb{C}^q$ and $\dim(H)=d=q^L$. The time evolution is discrete and is generated by a brickwork pattern of a (fixed) local unitary gate $V \in \mathcal{L}(\mathbb{C}^q \otimes \mathbb{C}^q)$. The evolution operator at time $t=0,1,\dots$ is given by $U_t= (U_oU_e)^t$, where $U_e= V^{\otimes L/2}$ and $U_o=\mathds{1}_1 \otimes V^{\otimes L/2 -1} \otimes \mathds{1}_L$.
\par Local quantum circuits can be conveniently depicted using graphical notation. Specifically, the local gate $V$ is represented by an orange box with lines indicating the input and output Hilbert spaces, from bottom to top. The Hermitian conjugate $V^\dagger$ is indicated by a ``shaded'' orange box,
\begin{equation}
V=\begin{tikzpicture}[baseline=(current  bounding  box.center), scale=.7] \Vgate{0}{0} \end{tikzpicture}, \quad V^\dagger = \begin{tikzpicture}[baseline=(current  bounding  box.center), scale=.7] \Vdaggate{0}{0} \end{tikzpicture}.
\end{equation}
We will also utilize a vectorized description for the operators. Given a basis $\{\ket{m}\}_{m=1}^{q}$ of the $\mathbb{C}^q$ one can define the map $\mathcal{L}(\mathbb{C}^q) \rightarrow \mathbb{C}^q\otimes \mathbb{C}^q$ given as
\begin{equation}
\ketbra{m}{n} \rightarrow \ket{m} \otimes \ket{n}.
\end{equation}  
and extended to $\mathcal{L}(\mathbb{C}^q)$ by linearity. The choice of basis for the mapping is not crucial and can be thought of as a fixed ``computational basis'' for the local Hilbert spaces. In the graphical notation, this vectorization procedure corresponds to a ``folding'' \cite{kosCircuitsSpaceTime2023}. For example, for the identity operator,
\begin{equation}
\mathds{1} = 
\begin{tikzpicture}[baseline=(current  bounding  box.center), scale=.7] \draw[thick] (0,0.7) -- (0,0);\end{tikzpicture} \rightarrow \ket{\mathds{1}}=\sum_m \ket{m} \otimes \ket{m} = 
\begin{tikzpicture}[baseline=(current  bounding  box.center), scale=.7]
    \draw[thick] (0, 0.7) -- (0, 0);
    
    \draw[thick,decorate,decoration={bent,aspect=0.3,amplitude=-1pt}] (0,0) -- (0.08,0.04);
    
    \draw[thick] (0.08, 0.04) -- (0.08, 0.74);
\end{tikzpicture} \coloneqq
\begin{tikzpicture}[baseline=(current  bounding  box.center), scale=.7]\draw[very thick] (0,0) -- (0,0.4);
\draw[thick, fill=white] (0,0) circle (0.1cm);  
\end{tikzpicture},
\end{equation}
where we use thick wires to denote the doubled local Hilbert space.

Given the fixed basis $\{\ket{m}\}_{m=1}^{d}$, we define a complex conjugation operation $*: \mathcal{L}(\mathbb{C}^q) \rightarrow \mathcal{L}(\mathbb{C}^q)$ as $\mel{m}{X^*}{n} \coloneqq \mel{m}{X}{n}^*$, $ X\in\mathcal{L}(\mathbb{C}^q)$. Then, the adjoint action of an operator $X$ is mapped to an operator in $\mathcal{L}(\mathcal{H}^{\otimes 2})$ in the vectorized picture,
\begin{equation}
\mathcal{X}[\bullet ]=X[\bullet ]X^\dagger \rightarrow X \otimes X^*.
\end{equation}
For the 2-quqit gate $V$, the ``folded'' operator is denoted as
\begin{equation} \label{folded}
V\otimes V^*  = 
\begin{tikzpicture}[baseline=(current  bounding  box.center), scale=.7]
\def\dx{0.15}
\def\dy{0.15}
\draw[ thick] (-4.25+\dx,0.5+\dy) -- (-3.25+\dx,-0.5+\dy);
\draw[ thick] (-4.25+\dx,-0.5+\dy) -- (-3.25+\dx,0.5+\dy);
\draw[ thick, fill=mygreen, rounded corners=2pt] (-4+\dx,0.25+\dy) rectangle (-3.5+\dx,-0.25+\dy);
\draw[ thick] (-4.25,0.5) -- (-3.25,-0.5);
\draw[ thick] (-4.25,-0.5) -- (-3.25,0.5);
\draw[ thick, fill=myorange, rounded corners=2pt] (-4,0.25) rectangle (-3.5,-0.25);
\draw[thick] (-3.75,0.15) -- (-3.75+0.15,0.15) -- (-3.75+0.15,0);
,
\end{tikzpicture} \coloneqq
\begin{tikzpicture}[baseline=(current  bounding  box.center), scale=.7] 
\Vfoldedgate{0}{0} 
\end{tikzpicture},
\end{equation}
where we use a green box to denote the complex conjugate $V^*$. It will also be convenient to define the complex conjugate of the ``folded'' operator,
\begin{equation} \label{folded_conj}
V^*\otimes V = 
\begin{tikzpicture}[baseline=(current  bounding  box.center), scale=.7]
\def\dx{0.15}
\def\dy{0.15}
\draw[ thick] (-4.25+\dx,0.5+\dy) -- (-3.25+\dx,-0.5+\dy);
\draw[ thick] (-4.25+\dx,-0.5+\dy) -- (-3.25+\dx,0.5+\dy);
\draw[ thick, fill=myorange, rounded corners=2pt] (-4+\dx,0.25+\dy) rectangle (-3.5+\dx,-0.25+\dy);
\draw[ thick] (-4.25,0.5) -- (-3.25,-0.5);
\draw[ thick] (-4.25,-0.5) -- (-3.25,0.5);
\draw[ thick, fill=mygreen, rounded corners=2pt] (-4,0.25) rectangle (-3.5,-0.25);
\draw[thick] (-3.75,0.15) -- (-3.75+0.15,0.15) -- (-3.75+0.15,0);
,
\end{tikzpicture} \coloneqq
\begin{tikzpicture}[baseline=(current  bounding  box.center), scale=.7] 
\Vfoldedstargate{0}{0} 
\end{tikzpicture}.
\end{equation}
\par With the above graphical conventions, the unitarity of $V$ can be equivalently expressed as
\begin{equation} \label{du_1}
\begin{split}
&V V^\dagger = \mathds{1} \otimes \mathds{1}, \quad 
\begin{tikzpicture}[baseline=(current  bounding  box.center), scale=.7]
\Vdaggate{0}{0}\Vgate{0}{1}
\end{tikzpicture}=
\begin{tikzpicture}[baseline=(current  bounding  box.center), scale=.7]
\def\dx{-0.25}
\draw[thick] (-0.5,1) -- (-0.25,0.75) -- (-0.25,0.25) -- (-0.5,0);
\draw[thick] (0.5+\dx,1) -- (0.25+\dx,0.75) -- (0.25+\dx,0.25) -- (0.5+\dx,0);
\end{tikzpicture}\, , \quad
\begin{tikzpicture}[baseline=(current  bounding  box.center), scale=.7]
\Vfoldedgate{0}{0}
\draw[thick, fill=white] (0-0.5,-0.5) circle (0.1cm); 
\draw[thick, fill=white] (0.5,-0.5) circle (0.1cm);
\end{tikzpicture}=
\begin{tikzpicture}[baseline=(current  bounding  box.center), scale=.7]
\Vfoldedstargate{0}{0}
\draw[thick, fill=white] (0-0.5,-0.5) circle (0.1cm); 
\draw[thick, fill=white] (0.5,-0.5) circle (0.1cm);
\end{tikzpicture}=
\begin{tikzpicture}[baseline=(current  bounding  box.center), scale=.7]
%
\draw[very thick] (0,0) -- (0,0.6);
\draw[thick, fill=white] (0,0) circle (0.1cm); 
%
\draw[very thick] (0.5,0) -- (0.5,0.6);
\draw[thick, fill=white] (0.5,0) circle (0.1cm); 
\end{tikzpicture}\\
&V^\dagger V = \mathds{1} \otimes \mathds{1}, \quad 
\begin{tikzpicture}[baseline=(current  bounding  box.center), scale=.7]
\Vgate{0}{0}\Vdaggate{0}{1}
\end{tikzpicture}=
\begin{tikzpicture}[baseline=(current  bounding  box.center), scale=.7]
\def\dx{-0.25}
\draw[thick] (-0.5,1) -- (-0.25,0.75) -- (-0.25,0.25) -- (-0.5,0);
\draw[thick] (0.5+\dx,1) -- (0.25+\dx,0.75) -- (0.25+\dx,0.25) -- (0.5+\dx,0);
\end{tikzpicture}\, , \quad
\begin{tikzpicture}[baseline=(current  bounding  box.center), scale=.7]
\Vfoldedgate{0}{0}
\draw[thick, fill=white] (0-0.5,0.5) circle (0.1cm); 
\draw[thick, fill=white] (0.5,0.5) circle (0.1cm);
\end{tikzpicture}=
\begin{tikzpicture}[baseline=(current  bounding  box.center), scale=.7]
\Vfoldedstargate{0}{0}
\draw[thick, fill=white] (0-0.5,0.5) circle (0.1cm); 
\draw[thick, fill=white] (0.5,0.5) circle (0.1cm);
\end{tikzpicture}=
\begin{tikzpicture}[baseline=(current  bounding  box.center), scale=.7]
%
\draw[very thick] (0,0) -- (0,0.6);
\draw[thick, fill=white] (0,0.6) circle (0.1cm); 
%
\draw[very thick] (0.5,0) -- (0.5,0.6);
\draw[thick, fill=white] (0.5,0.6) circle (0.1cm); 
\end{tikzpicture}.
\end{split}
\end{equation}
We will further assume that the local gate $V$ is dual-unitary (DU). The class of DU circuits provides a minimal model of locally interacting many-body systems, where many statistical and dynamical properties are analytically accessible \cite{bertiniExactCorrelationFunctions2019,bertiniExactSpectralForm2018,piroliExactDynamicsDualunitary2020}, despite being generically chaotic \cite{bertiniExactSpectralForm2018,bertiniRandomMatrixSpectral2021}. The dual-unitarity condition requires $V$ to remain unitary upon exchanging the role of space and time and is conveniently expressed in terms of the graphical notation as
\begin{equation} \label{du_2}
\begin{split}
&\begin{tikzpicture}[baseline=(current  bounding  box.center), scale=.7]
\Vdaggate{0}{0}\Vgate{1}{0}
\end{tikzpicture}=
\begin{tikzpicture}[baseline=(current  bounding  box.center), scale=.7]
\def\dx{0}
\draw[thick] (-0.5,1) -- (-0.25,0.75) -- (0.25+\dx,0.75) -- (0.5+\dx,1);
\draw[thick] (-0.5,0.25) -- (-0.25,0.5) --  (0.25+\dx,0.5) -- (0.5+\dx,0.25);
\end{tikzpicture}\, , \quad
\begin{tikzpicture}[baseline=(current  bounding  box.center), scale=.7]
\Vfoldedgate{0}{0}
\draw[thick, fill=white] (-0.5,-0.5) circle (0.1cm); 
\draw[thick, fill=white] (-0.5,0.5) circle (0.1cm);
\end{tikzpicture}=
\begin{tikzpicture}[baseline=(current  bounding  box.center), scale=.7]
\Vfoldedstargate{0}{0}
\draw[thick, fill=white] (-0.5,-0.5) circle (0.1cm); 
\draw[thick, fill=white] (-0.5,0.5) circle (0.1cm);
\end{tikzpicture}=
\begin{tikzpicture}[baseline=(current  bounding  box.center), scale=.7]
%
\draw[very thick] (0,0.25) -- (0.6,0.25);
\draw[thick, fill=white] (0,0.25) circle (0.1cm); 
%
\draw[very thick] (0,-0.25) -- (0.6,-0.25);
\draw[thick, fill=white] (0,-0.25) circle (0.1cm); 
\end{tikzpicture}\\
&\begin{tikzpicture}[baseline=(current  bounding  box.center), scale=.7]
\Vgate{0}{0}\Vdaggate{1}{0}
\end{tikzpicture}=
\begin{tikzpicture}[baseline=(current  bounding  box.center), scale=.7]
\def\dx{0}
\draw[thick] (-0.5,1) -- (-0.25,0.75) -- (0.25+\dx,0.75) -- (0.5+\dx,1);
\draw[thick] (-0.5,0.25) -- (-0.25,0.5) --  (0.25+\dx,0.5) -- (0.5+\dx,0.25);
\end{tikzpicture}\, , \quad
\begin{tikzpicture}[baseline=(current  bounding  box.center), scale=.7]
\Vfoldedgate{0}{0}
\draw[thick, fill=white] (0.5,0.5) circle (0.1cm); 
\draw[thick, fill=white] (0.5,-0.5) circle (0.1cm);
\end{tikzpicture}=
\begin{tikzpicture}[baseline=(current  bounding  box.center), scale=.7]
\Vfoldedstargate{0}{0}
\draw[thick, fill=white] (0.5,0.5) circle (0.1cm); 
\draw[thick, fill=white] (0.5,-0.5) circle (0.1cm);
\end{tikzpicture}=
\begin{tikzpicture}[baseline=(current  bounding  box.center), scale=.7]
%
\draw[very thick] (0,0.25) -- (0.6,0.25);
\draw[thick, fill=white] (0.6,0.25) circle (0.1cm); 
%
\draw[very thick] (0,-0.25) -- (0.6,-0.25);
\draw[thick, fill=white] (0.6,-0.25) circle (0.1cm); 
\end{tikzpicture}.
\end{split}
\end{equation}
We note that the dual-unitarity condition is equivalent to the condition that the operator entanglement of $V$ is maximal \cite{ratherConstructionLocalEquivalence2022a}, namely $E(V)=E(\mathsf{S}_{12})$ where $\mathsf{S}_{12}$ is the swap between the two sites the gate $V$ acts on. This means that $V$ scrambles information maximally between the two sites it acts on, but as noted in \cref{sec:opentp}, its operator space entangling power may range from zero to near-maximal. In particular, using \cref{opentp_sym} for the 2-site Hilbert space $\mathcal{H}_1 \otimes \mathcal{H}_2 \cong \mathbb{C}^q \otimes \mathbb{C}^q$, we have
\begin{equation}
E_p(\mathcal{V}) = \frac{E(\mathsf{S}_{12})}{E(\mathsf{S}_{12})}\left(2E(\mathsf{S}_{12})-\frac{E(V\mathsf{S}_{12})}{E(\mathsf{S}_{12})})\right).
\end{equation}
\par Our goal is to study the relation between the operator space entangling power of the local gate $V$ and properties of the full many-body dynamics it generates. In particular, we aim to investigate how the ability of $V$ to generate operator entanglement between neighboring sites correlates with the growth of the operator entanglement of ultra-local operators across a half-chain bipartition under the total evolution operator $U_t$. This latter quantity, referred to as local operator entanglement, has been studied in the context of dual unitary circuits for given ultra-local operators \cite{bertiniOperatorEntanglementLocal2022,bertiniOperatorEntanglementLocal2020b}. In order to avoid any operator dependence, we consider the average local operator entanglement over ultra-local operators on a single site. For simplicity, we choose the local operators to have support on the first site $\mathcal{H}_1$, and performing the average, we obtain, see \cref{app:loc},
\begin{widetext}
\begin{equation} \label{locopent_exact}
\begin{split}
&E_{\text{loc}}(t)\coloneqq \mathlarger{\mathbb{E}}_{X_1} E(\mathcal{U}_t(X_1 \otimes \mathds{1}_{\{2,\dots,L\}})) =\\
&=1-\frac{1}{d^2(q^2-1)} \left( \lVert \Tr_{11^\prime}({\mathcal{U}_t^\dagger}^{\otimes 2}(\mathsf{S}_{BB^\prime}) \rVert_2^2 + \lVert \Tr_{11^\prime}(\mathsf{S}_{11^\prime} \, {\mathcal{U}_t^\dagger}^{\otimes 2}(\mathsf{S}_{BB^\prime}) \rVert_2^2- \frac{2}{q} \langle \Tr_{11^\prime}({\mathcal{U}_t^\dagger}^{\otimes 2}(\mathsf{S}_{BB^\prime}),\Tr_{11^\prime}(\mathsf{S}_{11^\prime}\, {\mathcal{U}_t^\dagger}^{\otimes 2}(\mathsf{S}_{BB^\prime}) \rangle \right),
\end{split}
\end{equation}
\end{widetext}
where $A$ is the subsystem associated to the first half of the chain, $B$ the subsystem associated to the second half of the chain and the operator entanglement $E$ is taken with respect to the bipartition $A:B$. We should note that in this work we use the linear entropy as a measure of the operator entanglement, which is related to the R{\'e}nyi entropy utilized in other works \cite{bertiniOperatorEntanglementLocal2022,bertiniOperatorEntanglementLocal2020b} as $S_{lin}(\rho ) = 1- \exp(S_2(\rho ))$, where $S_2(\rho )$ is the 2-R{\'e}nyi entropy.
\par Due to the strict light-cone in local quantum circuits, $E_{\text{loc}}(t)=0\, \forall\,  t<t^{\star}\coloneqq L/2-\left\lfloor{L/4}\right\rfloor$. The exact dynamics of local operator entanglement in interacting systems for arbitrary times are significantly complex. Here, we provide an analytical result for the first non-trivial value $E_{\text{loc}}(t^{\star})$ for a DU quantum circuit, see \cref{app:tstar},
\begin{equation} \label{locopent}
\begin{split}
E_{\text{loc}}(t^{\star})=&1-\frac{1}{q^2-1}-\\
&-\frac{1}{q^2(q^2-1)}\left(\lVert P^{L/2} \rVert_2^2 - 2 \lVert M_-^{L/2} \rVert_2^2 \right),
\end{split}
\end{equation}
where\footnote{$M_-$ and $M_+$ in \cref{matrices,mp} contain the ``folded'' operator, while $P$ in \cref{matrices} contains one copy of the ``folded'' operator and one copy of the complex conjugate of the ``folded'' operator, see \cref{folded,folded_conj}.}
\begin{equation} \label{matrices}
M_-=\frac{1}{q}
\begin{tikzpicture}[baseline=(current  bounding  box.center), scale=.7]
\Vfoldedgate{0}{0}
\draw[thick, fill=white] (0.5,-0.5) circle (0.1cm);  
\draw[thick, fill=white] (-0.5,0.5) circle (0.1cm); 
\end{tikzpicture}\, , \quad
P=\frac{1}{q}
\begin{tikzpicture}[baseline=(current  bounding  box.center), scale=.7]
\draw[very thick] (-0.5, +0.5) -- (+0.25,-0.25);
\draw[very thick] (-0.5,-0.5) -- (+0.5,+0.5);
\draw[ thick, fill=myred, rounded corners=2pt] (-0.25,+0.25) rectangle (+0.25,-0.25);
\draw[thick] (0,+0.15) -- (+0.15,+0.15) -- (+0.15,0);
\def\dx{1.5}
\draw[very thick] (\dx-0.5, +0.5) -- (\dx+0.25,-0.25);
\draw[very thick] (\dx-0.5,-0.5) -- (\dx+0.5,+0.5);
\draw[ thick, fill=myblue, rounded corners=2pt] (\dx-0.25,+0.25) rectangle (\dx+0.25,-0.25);
\draw[thick] (\dx,+0.15) -- (\dx+0.15,+0.15) -- (\dx+0.15,0);
\draw[thick, fill=white] (-0.5,0.5) circle (0.1cm);  
\draw[thick, fill=white] (\dx-0.5,0.5) circle (0.1cm);
\draw[very thick,decorate,decoration={bent,aspect=0.3,amplitude=-15pt}] (0.25,-0.25) -- (\dx+0.25,-0.25);
\end{tikzpicture}.
\end{equation}
\begin{figure}
\centering
  \includegraphics[width=1\linewidth]{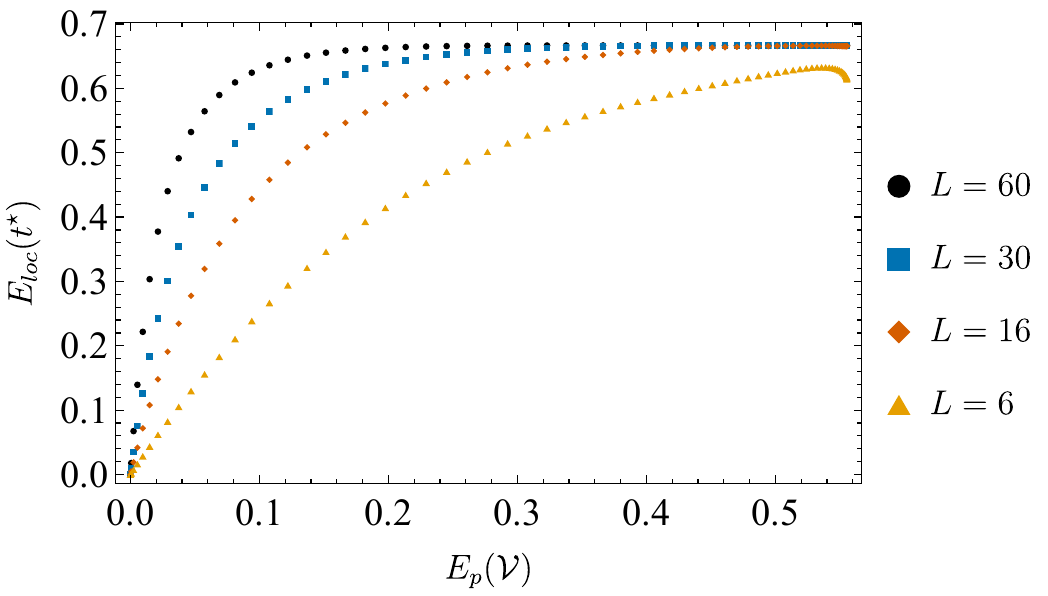}
\caption{Comparison of the behavior of $E_p(\mathcal{V})$ and $E_{\text{loc}}(t^{\star})$ for different system sizes $L$. For large system sizes $E_{\text{loc}}(t^*)$ saturates after some value $E_p^*$, while it grows monotonically with $E_p(\mathcal{V})$ for $E_p(\mathcal{V})<E_p^*$. }
\label{fig:locopent_comp}
\end{figure}
The expression in \cref{locopent} is efficiently computable, capturing the initial growth rate of local operator entanglement as the information light-cone reaches the bipartition boundary. The matrix $M_-$ in \cref{matrices} corresponds to one of the transfer matrices in lightcone coordinates for DU circuits due to its role in computing correlation functions \cite{bertiniExactCorrelationFunctions2019}. The other one is
\begin{equation} \label{mp}
M_+=\frac{1}{q}
\begin{tikzpicture}[baseline=(current  bounding  box.center), scale=.7]
\Vfoldedgate{0}{0}
\draw[thick, fill=white] (0.5,0.5) circle (0.1cm);  
\draw[thick, fill=white] (-0.5,-0.5) circle (0.1cm); 
\end{tikzpicture}.
\end{equation}
In addition, $M_-$, $M_+$ and $P$ satisfy the following relations,
\begin{align}
&\lVert M_- \rVert_2^2 = \lVert M_+ \rVert_2^2 =q^2 \left(1- E(V\mathsf{S}_{12}) \right),  \label{property_1}\\
&\lVert P \rVert_2^2 = q^2\lVert M_+ M_+^\dagger \rVert_2^2, \label{property_2}\\
&\lVert M_-\rVert_\infty = \lVert M_+\rVert_\infty = 1 \label{property_3}
\end{align}
%
\begin{figure*}
\centering
\begin{subfigure}{.5\textwidth}
  \centering
  \includegraphics[width=1\linewidth]{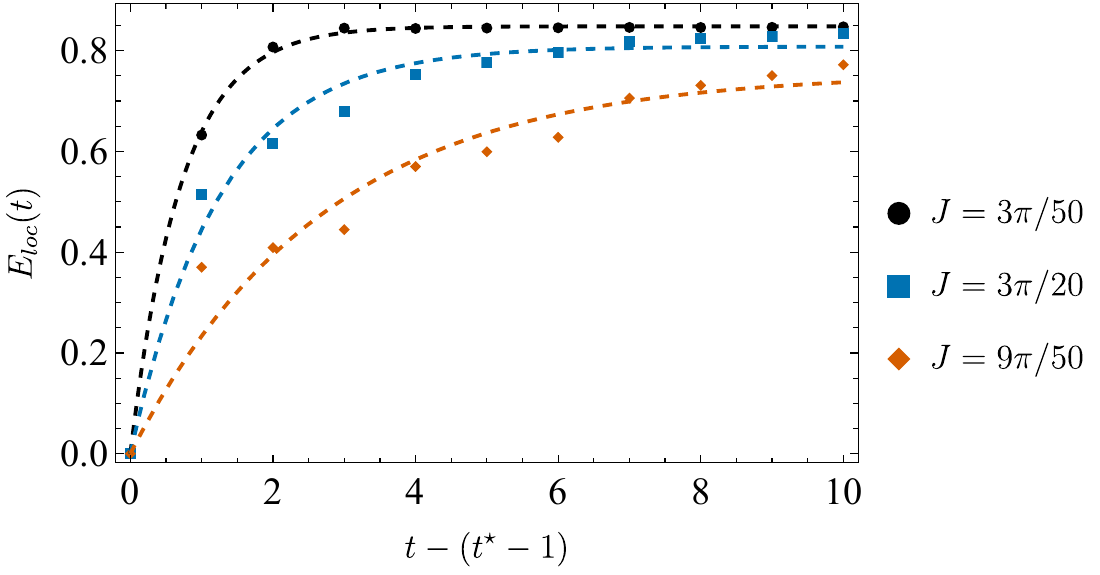}
 \caption{}\label{fig:locopentt6}
\end{subfigure}%
\begin{subfigure}{.5\textwidth}
  \centering
  \includegraphics[width=1\linewidth]{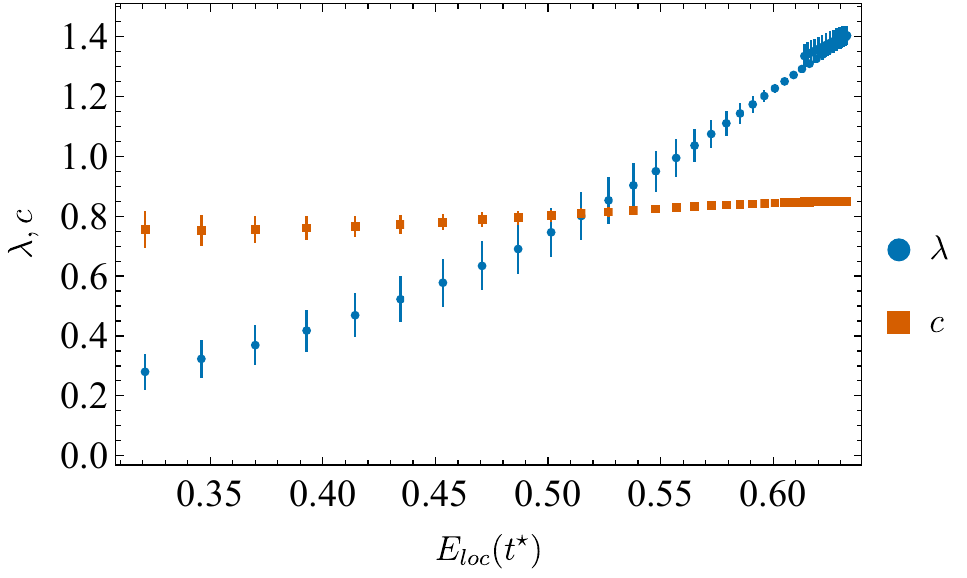}
  \caption{}\label{fig:parameters6}
\end{subfigure}
\caption{Exponential relaxation for the full time dynamics of the local operator entanglement for $L=6$. (a) Numerical simulation data alongside best-fit curves of $E_{\text{loc}}(t)=c\left(1-\exp[-\lambda (t-t^{\star}) ])\right)$ for various values of $J\in [0,\frac{3}{16}\pi]$. (b) The exponential growth rate $\lambda$ is positively correlated with $E_{\text{loc}}(t^{\star})$, while the equilibration value $c$ is virtually independent of the gate parameter $J\in [0,\frac{3}{16}\pi]$. }
\label{fig:locopent_time}
\end{figure*}
%
\par In the case of qubits, $q=2$, the structure of DU gates has been completely characterized \cite{bertiniExactCorrelationFunctions2019}:
\begin{equation} \label{du_char}
\begin{split}
V=&e^{i\phi } (u_+ \otimes u_-) \\
&\exp( -i\left(\frac{\pi}{4} \sigma_x \otimes \sigma_x + \frac{\pi}{4} \sigma_y \otimes \sigma_y + J \sigma_z \otimes \sigma_z \right) )\\
&(v_+ \otimes v_-),
\end{split}
\end{equation}
where $\phi , J \in \mathbb{R}$, $u_\pm, v_\pm \in SU(2)$. Also, under the gauge transformation
\begin{equation} \label{gauge}
V \rightarrow (v \otimes u) V (u^\dagger \otimes v^\dagger),
\end{equation}
we have  \cite{bertiniOperatorEntanglementLocal2022}
\begin{equation}
\begin{split}
&\mathcal{U}_t(X_1 \otimes \mathds{1}_{\{2,\dots,L\}}) \rightarrow \\
&\hspace{20pt}\rightarrow(v\otimes u)^{\otimes L/2} \mathcal{U}_t(u^\dagger X_1 u \otimes \mathds{1}_{\{2,\dots,L\}}) (v^\dagger \otimes u^\dagger)^{\otimes L/2}.
\end{split}
\end{equation}
Due to the invariance of the operator entanglement under local unitaries,
\begin{equation}
\begin{split}
&E\left((v\otimes u)^{\otimes L/2} \, \mathcal{U}_t\left(u^\dagger X_1 u \otimes \mathds{1}_{\{2,\dots,L\}}\right) (v^\dagger \otimes u^\dagger)^{\otimes L/2}\right)= \\
&\hspace{20pt}=E\left(\mathcal{U}_t\left(u^\dagger X_1 u \otimes \mathds{1}_{\{2,\dots,L\}}\right)\right)
\end{split}
\end{equation}
and due to the left and right invariance of the Haar measure, 
\begin{equation}
\begin{split}
&\mathbb{E}_{X_1} E\left(\mathcal{U}_t\left(u^\dagger X_1 u \otimes \mathds{1}_{\{2,\dots,L\}}\right)\right)= \\
&\hspace{20pt} =\mathbb{E}_{X_1} E\left(\mathcal{U}_t\left(X_1 \otimes \mathds{1}_{\{2,\dots,L\}}\right)\right).
\end{split}
\end{equation}
Thus, $E_{\text{loc}}(t)$ is invariant under the gauge transformation \cref{gauge} and without loss of generality we can fix $\phi =0 , u_+= u_- =\mathds{1}$ in \cref{du_char}.
\par $E_p(\mathcal{V})$ does not depend on the local unitaries and is a function of $J$:
\begin{equation} \label{opentp_J}
E_p(\mathcal{V}) = \frac{\cos^2(2J)}{9} \left(7-2\cos(4J) \right).
\end{equation}
From \cref{opentp_J}, we see that $E_p(\mathcal{V})$ is a periodic function of $J$ with period $\pi/2$, and symmetric around $\pi/4$. For $J=\pi/4$ the interaction term of the local gate $V$ is $\exp( -i\left(\frac{\pi}{4} \sigma_x \otimes \sigma_x + \frac{\pi}{4} \sigma_y \otimes \sigma_y + \frac{\pi}{4} \sigma_z \otimes \sigma_z \right) )= \mathsf{S}_{12}$ and the dynamics consist of local unitaries and swaps leading to $E_p(\mathcal{V})=0$, while as $J$ decreases from $\pi /4$ to $0$, $E_p(\mathcal{V})$ grows monotonically.

\par Using \cref{locopent}, we numerically compute $E_{\text{loc}}(t^{\star})$ also as a function of $J$ for a fixed choice of $v_+,v_-$\footnote{See \cref{appendix_du} for details about the gate choices in the numerical simulations.}. We compare the behavior of $E_{\text{loc}}(t^{\star})$ with $E_p(\mathcal{V})$ in \cref{fig:locopent_comp}. We observe that $E_{\text{loc}}(t^{\star})$ generally increases with $E_p(\mathcal{V})$. This is most clear for larger system sizes, where we observe two distinct regimes: (i) for $E_p(\mathcal{V} ) > E_p^*$, $E_{\text{loc}}(t^{\star})$ saturates to a maximal value, (ii) for $E_p(\mathcal{V} ) < E_p^*$, $E_{\text{loc}}(t^{\star})$ grows monotonically with $E_p(\mathcal{V})$. The size of the ``maximal growth'' region (i) is controlled by $E_p^*$ and increases with the system size. We note that the behavior of $E_{\text{loc}}(t^{\star})$ as a function of $E_p(\mathcal{V})$---which depends monotonically on the gate parameter $J$---in \cref{fig:locopent_comp} is congruent to an asymptotic result for the growth of local operator entanglement in chaotic DU circuits \cite{bertiniOperatorEntanglementLocal2022}.

\par A natural question is how much information $E_{\text{loc}}(t^{\star})$ carries about the full time dynamics $E_{\text{loc}}(t)$. In the absence of local conservation laws, DU circuits are generically chaotic and the local operator entanglement is expected to have an exponential relaxation form $E_{\text{loc}}(t)=c\left(1-\exp[-\lambda (t-t^{\star})]\right)$\footnote{This exponential relaxation for the linear entropy corresponds to the linear growth of the 2-R{\'e}nyi entropy in Ref. \cite{bertiniOperatorEntanglementLocal2022}.}. Simulating the exact DU circuit dynamics for $L=6$, we evaluate \cref{locopent_exact} and fit the numerical data to the curve $E_{\text{loc}}(t)=c\left(1-\exp[-\lambda (t-t^{\star})]\right)$  to obtain the coefficients $c,\lambda$ for various values of the gate parameter $J\in [0,\frac{3}{16}\pi]$, see \cref{fig:locopent_time}.
We select values of $J$ deviating from $\frac{\pi}{4}$, as the timescale $\lambda^{-1}$ diverges for $J=\frac{\pi}{4}$ and the evolution time required to accurately capture the form of the dynamics as $J$ approaches $\frac{\pi}{4}$ exceeds the capabilities of exact simulations. We observe that $c$, which determines the equilibration value, is effectively independent of $J$, and $\lambda$ is positively correlated with $E_{\text{loc}}(t^{\star})$. This means that the exponential growth rate of the local operator entanglement for chaotic DU circuits is well-correlated with the value at $t^{\star}$ for a given system size $L$.

\section{Conclusion}\label{sec:conclusion}
We introduced a measure, the operator space entangling power $E_p$, that quantifies the ability of a unitary channel to generate operator entanglement across a given bipartition. This quantity corresponds to the on-average operator entanglement generation for initially product operators and is analytically expressed in terms of mutual averaged non-commutativities of the subsystem operator algebras. The latter in turn are directly related to the subsystem entropy variations in the two subsystems under the unitary evolution. The operator space entangling power  is upper-bounded by the typical value of the operator entanglement, which intuitively corresponds to evolutions that map Haar random product operators to sufficiently uniformly distributed operators in the full operator space. These evolutions emulate the scrambling properties of Haar random unitaries in terms of \cref{max_condition}.

\par For unitary evolutions generated by a Hamiltonian, the short-time growth of the operator space entangling power is controlled by the Gaussian scrambling rate $\tau_s^{-1}$, which was introduced in Ref. \cite{zanardiOperationalQuantumMereology2024} and characterizes the short-time properties of quantum scrambling between the subsystems in terms of a bipartite of out-of-time-order-correlator. Numerical simulations of quantum many-body Hamiltonian models show that the full-time dynamics of the operator space entangling power exhibit a distinct behavior for localized models, where operator entanglement generation is suppressed even at late times. Nevertheless, the operator space entangling power is not effective in detecting integrability, despite the associated entropy ``exchange'' between the subsystems. This is attributed to an intrinsic property in the definition of the operator space entangling power, which renders it largely insensitive to any underlying locality structure beyond the given bipartition.
\par Finally, we have investigated the role of the operator space entangling power in the context of dual-unitary local quantum circuits. Specifically, we have examined the correlation between the operator space entangling power of the circuit-generating local gate $V$ and a property of the full system dynamics represented by the growth of the local operator entanglement, namely the operator entanglement of initially local operators. We obtained an analytical result for the first non-trivial value of the local operator entanglement and observed two distinct regimes depending on the operator space entangling power $E_p(\mathcal{V})$ of the local gate: for sufficiently large values of $E_p(\mathcal{V})$, the growth rate of local operator entanglement saturates, while for small values of $E_p(\mathcal{V})$, the growth rate of local operator entanglement monotonically increases with $E_p(\mathcal{V})$.
\par This work contributes to the ongoing effort of elucidating properties of quantum evolution through the analysis of operator space dynamics. Several paths for future research naturally emerge as a result. The gradient ascent search in \cref{subsec:bound} provides evidence that the upper bound in \cref{reduce_bound} is achievable, making it tantalizing to consider whether there is a general construction for the unitaries $U_{\star}$ that saturate it. Additionally, it is worthwhile to examine the role of the operator space entangling power of the local gate $V$ in dual unitary circuits with local conservation laws, as the dynamics of local operator entanglement in these circuits exhibit a distinct behavior compared to chaotic dual unitary models \cite{bertiniOperatorEntanglementLocal2020b}. Finally, investigating the relations of the operator space entangling power of quantum gates with other characteristic properties, such as the nonstabilizing power \cite{leoneStabilizerRenyiEntropy2022} or Pauli-weight dynamics \cite{schusterOperatorGrowthOpen2023,garciaResourceTheoryQuantum2023}, may prove to be a fruitful task.

\begin{acknowledgments}
FA acknowledges valuable comments from Namit Anand on the paper and insightful discussions with Georgios Styliaris on dual unitary circuits. FA and PZ acknowledge useful comments from Arul Lakshminarayan about an earlier version of the draft. FA and PZ acknowledge partial support from the NSF award PHY-2310227. FA acknowledges partial support from the Gerondelis foundation. This research was (partially) sponsored by the Army Research Office and 
was accomplished under Grant Number W911NF-20-1-0075.  
The views and conclusions contained in this document are those of 
the authors and should not be interpreted as representing the official 
policies, either expressed or implied, of the Army Research Office
 or the U.S. Government. The U.S. Government is authorized to reproduce and 
distribute reprints for Government purposes notwithstanding any copyright
 notation herein.
\end{acknowledgments}

\bibliographystyle{apsrev4-2}
\bibliography{main.bib}

\onecolumngrid
\newpage
\appendix
\section{Analytical Derivations}\label{appendix_analytical}
\subsection{Proof of \cref{opentp_general,entropy_2}} \label{app:proofs_1}
Using \cref{opent}---with $A$ substituted by $B$---and \cref{opentp_def}, we have
\begin{equation} \label{opentp_explicit}
\begin{split}
E_p(\mathcal{U}) &= 1-\frac{1}{d^2}{\mathlarger{\mathbb{E}}}_{X_A,Y_B} \Tr( \mathsf{S}_{BB^\prime} (\mathcal{U}(X_A\otimes Y_B))^{\otimes 2} \mathsf{S}_{BB^\prime} (\mathcal{U}(X_A^\dagger \otimes Y_B^\dagger))^{\otimes 2}) \\
&=1-\frac{1}{d^2}{\mathlarger{\mathbb{E}}}_{X_A,Y_B} \Tr( {\mathcal{U}^\dagger}^{\otimes 2} (\mathsf{S}_{BB^\prime}) \, X_A \otimes Y_B \otimes X_{A^\prime} \otimes Y_{B^\prime} \, \mathcal{U}^{\otimes 2}(\mathsf{S}_{BB^\prime})\, X_A^\dagger \otimes Y_B^\dagger \otimes X^\dagger_{A^\prime} \otimes Y^\dagger_{B^\prime}).
\end{split}
\end{equation}
Schur-Weyl duality \cite{goodmanSymmetryRepresentationsInvariants2009} asserts that the commutant of the algebra generated by $\{V^{\otimes k} \vert V\in \mathcal{L}(\mathcal{H_\chi}), VV^\dagger =\mathds{1}_{\chi}\}$ is the group algebra of the symmetric group $\mathbf{C}S_k$. Also, $\mathbb{P} [\bullet ] \coloneqq \mathbb{E}_V V^{\otimes k} [\bullet ] {V^\dagger}^{\otimes k}$ is a projector onto this commutant, which for $k=2$ is spanned by the orthonormal basis $\left\{\frac{\mathds{1}_{\chi\chi^\prime}+\mathsf{S}_{\chi \chi^\prime}}{\sqrt{2d_\chi(d_\chi+1)}},\frac{\mathds{1}_{\chi\chi^\prime}-\mathsf{S}_{\chi\chi^\prime}}{\sqrt{2d_\chi(d_\chi-1)}}\right\}$. As a consequence,
\begin{equation} \label{schur}
{\mathlarger{\mathbb{E}}}_V V^{\otimes 2} [\bullet ] {V^\dagger}^{\otimes 2} = \sum_{\eta = \pm 1} \frac{\mathds{1}_{\chi\chi^\prime}+\eta \mathsf{S}_{\chi\chi^\prime}}{2d_\chi (d_\chi +\eta )} \langle \mathds{1}_{\chi\chi^\prime}+\eta \mathsf{S}_{\chi\chi^\prime}, \bullet \rangle .
\end{equation}
Applying \cref{schur} in \cref{opentp_explicit} for the two subsystems $\chi = A,B$, we get
\begin{equation} \label{opentp_derivation}
\begin{split}
E_p(\mathcal{U})&= 1- \sum_{\eta , \eta^\prime = \pm 1} \frac{\left\langle (\mathds{1}_{AA^\prime} + \eta \mathsf{S}_{AA^\prime}) \otimes (\mathds{1}_{BB^\prime} + \eta^\prime \mathsf{S}_{BB^\prime}), {\mathcal{U}^\dagger}^{\otimes 2}(\mathsf{S}_{BB^\prime}) \right\rangle  \Tr( {\mathcal{U}^\dagger}^{\otimes 2}(\mathsf{S}_{BB^\prime})\, (\mathds{1}_{AA^\prime} + \eta \mathsf{S}_{AA^\prime}) \otimes (\mathds{1}_{BB^\prime} + \eta^\prime \mathsf{S}_{BB^\prime}))}{4d^3(d_A+\eta )(d_B + \eta^\prime)}=\\
&=1-\sum_{\eta , \eta^\prime = \pm 1} \frac{\left[\Tr( {\mathcal{U}^\dagger}^{\otimes 2}(\mathsf{S}_{BB^\prime})\, (\mathds{1}_{AA^\prime} + \eta \mathsf{S}_{AA^\prime}) \otimes (\mathds{1}_{BB^\prime} + \eta^\prime \mathsf{S}_{BB^\prime}))\right]^2}{4d^3(d_A+\eta )(d_B + \eta^\prime)}=\\
&=\frac{d_B^2}{d_A^2} \frac{d_A^2-1}{d_B^2-1} \left[1-\left(1-\frac{S(\mathcal{U}(\mathcal{A}):\mathcal{B})}{S(\mathcal{A}:\mathcal{A})}\right)^2- \frac{d_A^2}{d_B^2} \left( 1-\frac{S(\mathcal{U}(\mathcal{A}):\mathcal{A})}{S(\mathcal{A}:\mathcal{A})}\right)^2-\frac{2}{d_B^2}\frac{S(\mathcal{U}(\mathcal{A}):\mathcal{B})}{S(\mathcal{A}:\mathcal{A})}\frac{S(\mathcal{U}(\mathcal{A}):\mathcal{A})}{S(\mathcal{A}:\mathcal{A})}\right],
\end{split}
\end{equation}
where
\begin{align} 
&S(\mathcal{U}(\mathcal{A}):\mathcal{B})=1-\frac{1}{d} \Tr(\mathsf{S} \, {\mathcal{U}}^{\otimes 2}\left(\frac{\mathsf{S}_{AA^\prime}}{d_A}\right) \frac{\mathsf{S}_{BB^\prime}}{d_B} ) =1-\frac{1}{d^2} \Tr(\mathsf{S}_{BB^\prime} {\mathcal{U}^\dagger}^{\otimes 2}(\mathsf{S}_{BB^\prime})),\label{man1} \\
&S(\mathcal{U}(\mathcal{A}):\mathcal{A})=1-\frac{1}{d} \Tr(\mathsf{S} \, {\mathcal{U}}^{\otimes 2}\left(\frac{\mathsf{S}_{AA^\prime}}{d_A}\right) \frac{\mathsf{S}_{AA^\prime}}{d_A})=1-\frac{1}{dd_A^2} \Tr(\mathsf{S}_{AA^\prime} {\mathcal{U}^\dagger}^{\otimes 2}(\mathsf{S}_{BB^\prime})),\label{man2}
\end{align}
which proves \cref{opentp_general}.
\par \cref{entropy_2} follows in a similar way to \cref{entropy_1}, which is Theorem 8 of Ref. \cite{styliarisInformationScramblingBipartitions2021}. The basic ingredient is rewriting $\mathsf{S}_{AA^\prime}$ using an average over pure states,
\begin{equation} \label{swap_pure}
\mathbb{E}_{\ket{\psi}} \ketbra{\psi}{\psi}^{\otimes 2} = (\mathds{1}_{AA^\prime} + \mathsf{S}_{AA^\prime})/(d_A(d_A+1)),
\end{equation}
where $\mathbb{E}_{\ket{\psi}}$ denotes the average over Haar random pure states on $\mathcal{H}_A$. Using \cref{swap_pure} in \cref{man1} we get
\begin{equation} \label{man_deriv_1}
\begin{split}
S(\mathcal{U}(\mathcal{A}):\mathcal{B})&= 1-\frac{1}{d^2} \left( -\Tr(\mathsf{S}_{AA^\prime} \otimes \mathds{1}_{BB^\prime}) + d_A (d_A+1) \mathbb{E}_{\ket{\psi}} \Tr( \mathsf{S}_{AA^\prime} \big( \, \mathcal{U} \left(\ketbra{\psi}{\psi} \otimes \mathds{1}_B \right)\big)^{\otimes 2} )\right) =\\
&= \frac{d_A+1}{d_A} \left( 1 - \Tr_A \left( \left[ \Tr_B \left( \mathcal{U} \left( \ketbra{\psi}{\psi}\otimes \frac{\mathds{1}_B}{d_B} \right) \right) \right]^2 \right) \right),
\end{split}
\end{equation}
which is \cref{entropy_1} since $S_{lin}(\rho_{AB} ) = 1- \Tr_A( \left[ \Tr_B(\rho_{AB})\right])$. To get to the second line of \cref{man_deriv_1} we also used the identity $\Tr( \mathsf{S}_{AA^\prime} X^{\otimes 2})=\Tr_A( \left[\Tr_B(X)\right]^2)$. Similarly, for \cref{man2},
\begin{equation} \label{man_deriv_2}
S(\mathcal{U}(\mathcal{A}):\mathcal{A})= \frac{d_A+1}{d_A} \left( 1 - \frac{d_B}{d_A} \Tr_B \left( \left[ \Tr_A \left( \mathcal{U} \left( \ketbra{\psi}{\psi} \otimes \frac{\mathds{1}_B}{d_B} \right) \right) \right]^2 \right) \right),
\end{equation}
which is \cref{entropy_2}.
\subsection{Proof of $S(\mathcal{U}_t(\mathcal{A}):\mathcal{A}) = 1 -\frac{1}{d_A^2} + O(t^3)$} \label{app:short}
Consider the series expansion $U_t=\exp(iHt)=\mathds{1}+iHt-\frac{H^2t^2}{2} + O(t^3)$. Then, for ${\mathcal{U}_t^\dagger}^{\otimes 2}[ \bullet ] = U_t^\dagger \otimes U_t^\dagger [\bullet ] U_t \otimes U_t$ we have
\begin{equation} \label{expansion}
\begin{split}
{\mathcal{U}_t^\dagger}^{\otimes 2}[ \bullet ] = &\mathds{1} + it \left[ (\mathds{1} \otimes H + H \otimes \mathds{1}), \bullet \right] +\\
&+ t^2 \left( (\mathds{1} \otimes H + H \otimes \mathds{1}) \bullet (\mathds{1} \otimes H + H \otimes \mathds{1}) - \left\{ \frac{(\mathds{1}\otimes H^2 + H^2 \otimes \mathds{1})}{2} + H \otimes H , \bullet \right \} \right) + O(t^3),
\end{split}
\end{equation}
where the brackets $[\cdot,\cdot]$ denote the commutator and the curly brackets $\{\cdot,\cdot\}$ the anti-commutator. On account of the identities 
\begin{equation}
\begin{split}
&\Tr(\mathsf{S})=d,\\
&\Tr( \mathsf{S}_{AA^\prime} [\mathds{1} \otimes H + H \otimes \mathds{1} , \mathsf{S}_{BB^\prime} ]) =0,\\
&\Tr( \mathsf{S}_{AA^\prime} \, \mathds{1}\otimes H \, \mathsf{S}_{BB^\prime} \, \mathds{1} \otimes H)= \Tr( \mathsf{S}_{AA^\prime}\, H \otimes \mathds{1} \, \mathsf{S}_{BB^\prime} \, \mathds{1} \otimes H) = \lVert H \rVert_2^2,\\
&\Tr(\mathsf{S} \, \mathds{1}\otimes H^2) = \Tr(\mathsf{S} \, H \otimes H) = \lVert H \rVert_2^2,
\end{split}
\end{equation}
and using \cref{expansion} in \cref{man2}, it follows that both the linear and quadratic term in $S(\mathcal{U}_t(\mathcal{A}):\mathcal{A})$ are zero, while the constant term is $1-\frac{1}{d_A^2}$.

\subsection{Proof of \cref{opentp_bound}} \label{app:max}
This follows simply from the functional form \cref{opentp_general}. Treating $S(\mathcal{U}(\mathcal{A}):\mathcal{B})\equiv x$, $S(\mathcal{U}(\mathcal{A}):\mathcal{A})\equiv y$ as independent variables with $x,y \in[0, 1-\frac{1}{d_A^2}]$, $d_A \leq d_B$,
\begin{equation}
\begin{split}
&\frac{\partial E_p(\mathcal{U})}{\partial x} = 2N(d_A,d_B) \left(1-x-\frac{y}{d_B^2} \right),\\
&\frac{\partial E_p(\mathcal{U})}{\partial y} = 2N(d_A,d_B) \left(\frac{d_A^2}{d_B^2} (1-y) - \frac{x}{d_B^2}\right).
\end{split}
\end{equation}
Then, the condition $\frac{\partial E_p(\mathcal{U})}{\partial x}=\frac{\partial E_p(\mathcal{U})}{\partial y}=0$ yields $x_\star=\frac{(d_A^2-1)(d_B^2-1)}{d^2-1}$, $y_\star=\frac{d_B^2 (d_A^2-1)^2}{d_A^2(d^2-1)}$. Since 
\begin{equation*}
\begin{split}
&\frac{\partial^2 E_p(\mathcal{U})}{\partial x^2}\frac{\partial^2 E_p(\mathcal{U})}{\partial y^2}-\left(\frac{\partial^2 E_p(\mathcal{U})}{\partial x\partial y}\right)^2 = \frac{4N(d_A,d_B)^2}{d_B^4} (d^2-1) >0,\\
&\frac{\partial^2 E_p(\mathcal{U})}{\partial x^2}=-2N(d_A,d_B) <0,
\end{split}
\end{equation*}
this corresponds to the global maximum $\frac{(d_A^2-1)(d_B^2-1)}{d^2-1}$ of the function at $x_\star,y_\star$.
\par Using \cref{schur} it is straightforward to show that
\begin{equation}
\begin{split}
&\mathbb{E}_U S(\mathcal{U}(\mathcal{A}):\mathcal{B}) = \frac{(d_A^2-1)(d_B^2-1)}{d^2-1},\\ 
&\mathbb{E}_U S(\mathcal{U}(\mathcal{A}):\mathcal{A}) = \frac{d_B^2}{d_A^2} \frac{(d_A^2-1)^2}{d^2-1},
\end{split}
\end{equation}
which is \cref{max_condition} in the main text.

\subsection{Proof of \cref{locopent_exact}} \label{app:loc}
Using \cref{opent}---with $A$ substituted by $B$---, we have
\begin{equation}
\begin{split}
E_{\text{loc}}(t)&=1-\frac{1}{d^2} {\mathlarger{\mathbb{E}}}_{X_1} \Tr( \mathsf{S}_{BB^\prime} \left(\mathcal{U}_t(X_1 \otimes \mathds{1}_{\{2,\dots ,L\}})\right)^{\otimes 2} \mathsf{S}_{BB^\prime} \left(\mathcal{U}_t(X_1^\dagger \otimes \mathds{1}_{\{2,\dots ,L\}})\right)^{\otimes 2})\\
&=1-\frac{1}{d^2} {\mathlarger{\mathbb{E}}}_{X_1} \Tr( {\mathcal{U}^\dagger}^{\otimes 2} (\mathsf{S}_{BB^\prime}) \, X_1 \otimes \mathds{1}_{\{2,\dots ,L\}}\otimes X_{1^\prime} \otimes \mathds{1}_{\{2,\dots ,L\}} \, {\mathcal{U}^\dagger}^{\otimes 2} (\mathsf{S}_{BB^\prime})\, X_1^\dagger \otimes \mathds{1}_{\{2,\dots ,L\}}\otimes X_{1^\prime}^\dagger \otimes \mathds{1}_{\{2,\dots ,L\}}).
\end{split}
\end{equation}
Using \cref{schur} for the first quqit, we get
\begin{equation} \label{opentloc_proof}
\begin{split}
    E_{\text{loc}}(t)&=1-\frac{1}{d^2}\sum_{\eta = \pm 1} \frac{ \Tr( {\mathcal{U}^\dagger_t}^{\otimes 2}(\mathsf{S}_{BB^\prime}) \left[ (\mathds{1}_{11^\prime} + \eta \mathsf{S}_{11^\prime})\otimes\Tr_{11^\prime}((\mathds{1}_{11^\prime} + \eta \mathsf{S}_{11^\prime})\, {\mathcal{U}^\dagger_t}^{\otimes 2}(\mathsf{S}_{BB^\prime}))\right])}{2q(q+\eta)}=\\
    &= 1-\frac{1}{d^2}\sum_{\eta = \pm 1} \frac{ \left\lVert  \Tr_{11^\prime}((\mathds{1}_{11^\prime} + \eta \mathsf{S}_{11^\prime})\, {\mathcal{U}^\dagger_t}^{\otimes 2}(\mathsf{S}_{BB^\prime})) \right\rVert_2^2}{2q(q+\eta)} =\\
    &= 1-\frac{1}{d^2(q^2-1)} \left( \lVert \Tr_{11^\prime}({\mathcal{U}_t^\dagger}^{\otimes 2}(\mathsf{S}_{BB^\prime}) \rVert_2^2 + \lVert \Tr_{11^\prime}(\mathsf{S}_{11^\prime} {\mathcal{U}_t^\dagger}^{\otimes 2}(\mathsf{S}_{BB^\prime}) \rVert_2^2 - \frac{2}{q} \langle \Tr_{11^\prime}({\mathcal{U}_t^\dagger}^{\otimes 2}(\mathsf{S}_{BB^\prime}),\Tr_{11^\prime}(\mathsf{S}_{11^\prime} {\mathcal{U}_t^\dagger}^{\otimes 2}(\mathsf{S}_{BB^\prime}) \rangle \right),
\end{split}
\end{equation}
which is \cref{locopent_exact}.
\subsection{Derivation of \cref{locopent}} \label{app:tstar}
This follows from \cref{locopent_exact} for $t^{\star}=\frac{L}{2} - \left\lfloor \frac{L}{4} \right \rfloor$ using the dual-unitarity of $V$ \cref{du_1,du_2}. Let us show this using the graphical notation for $L=6$.
\begin{equation} \label{pic_1}
\begin{split}
    \Tr_{11^\prime}({\mathcal{U}_{t^{\star}}^\dagger}^{\otimes 2}(\mathsf{S}_{BB^\prime})) &= \begin{tikzpicture}[baseline=(current  bounding  box.center), scale=.7]
    \def\ds{0.5}
    \def\d{0.25}
    \def\dx{7}
    \draw[very thick] (0,0)--(0,-2*\ds);
    \draw[thick, fill=white] (0,0) circle (0.1cm);
    \Vfoldedgate{3*\ds}{-\ds}
    \draw[thick, fill=white] (2*\ds,0) circle (0.1cm);
    \draw[thick, fill=white] (4*\ds,0) circle (0.1cm);
	\Vfoldedgate{7*\ds}{-\ds}
	\draw[very thick] (7*\ds-\ds,-\ds+\ds)--(7*\ds-\ds,-\ds+\ds+3*\d);  
	\draw[very thick] (9*\ds-\ds,-\ds+\ds)--(9*\ds-\ds,-\ds+\ds+2*\d);
	\draw[very thick] (5,\d)--(5,-2*\ds);
	\Vfoldedgate{\ds}{-3*\ds}
	\Vfoldedgate{\ds+4*\ds}{-3*\ds}
	\Vfoldedgate{\ds+8*\ds}{-3*\ds}
	\Vfoldedgate{3*\ds}{-\ds-4*\ds}
	\Vfoldedgate{7*\ds}{-\ds-4*\ds}
	\draw[very thick](0,-4*\ds)--(0,-6*\ds);
	\draw[very thick](5,-4*\ds)--(5,-6*\ds);
	\Vfoldedgate{\ds}{-7*\ds}
	\Vfoldedgate{\ds+4*\ds}{-7*\ds}
	\Vfoldedgate{\ds+8*\ds}{-7*\ds}
	\draw[thick, fill=white] (0,-8*\ds) circle (0.1cm);
	%
	%
    \draw[very thick] (0+\dx,0)--(0+\dx,-2*\ds);
    \draw[thick, fill=white] (0+\dx,0) circle (0.1cm);
    \Vfoldedstargate{3*\ds+\dx}{-\ds}
    \draw[thick, fill=white] (2*\ds+\dx,0) circle (0.1cm);
    \draw[thick, fill=white] (4*\ds+\dx,0) circle (0.1cm);
	\Vfoldedstargate{7*\ds+\dx}{-\ds}
	\draw[very thick] (7*\ds-\ds+\dx,-\ds+\ds)--(7*\ds-\ds+\dx,-\ds+\ds+3*\d);  
	\draw[very thick] (9*\ds-\ds+\dx,-\ds+\ds)--(9*\ds-\ds+\dx,-\ds+\ds+2*\d);
	\draw[very thick] (5+\dx,\d)--(5+\dx,-2*\ds);
	\Vfoldedstargate{\ds+\dx}{-3*\ds}
	\Vfoldedstargate{\ds+4*\ds+\dx}{-3*\ds}
	\Vfoldedstargate{\ds+8*\ds+\dx}{-3*\ds}
	\Vfoldedstargate{3*\ds+\dx}{-\ds-4*\ds}
	\Vfoldedstargate{7*\ds+\dx}{-\ds-4*\ds}
	\draw[very thick](0+\dx,-4*\ds)--(0+\dx,-6*\ds);
	\draw[very thick](5+\dx,-4*\ds)--(5+\dx,-6*\ds);
	\Vfoldedstargate{\ds+\dx}{-7*\ds}
	\Vfoldedstargate{\ds+4*\ds+\dx}{-7*\ds}
	\Vfoldedstargate{\ds+8*\ds+\dx}{-7*\ds}
	\draw[thick, fill=white] (0+\dx,-8*\ds) circle (0.1cm);
	%
	%
	\draw[very thick] (7*\ds-\ds,-\ds+\ds+3*\d) -- (7*\ds-\ds+\dx,-\ds+\ds+3*\d);
	\draw[very thick] (9*\ds-\ds,-\ds+\ds+2*\d) -- (9*\ds-\ds+\dx,-\ds+\ds+2*\d);
	\draw[very thick] (11*\ds-\ds,-\ds+\ds+\d) -- (11*\ds-\ds+\dx,-\ds+\ds+\d);
    \end{tikzpicture}=\\
    %
    %
    %
    %
    %
    %
    &=\begin{tikzpicture}[baseline=(current  bounding  box.center), scale=.7]
    \def\ds{0.5}
    \def\d{0.25}
    \def\dx{7}
    \draw[very thick] (0,0)--(0,-3);
    \draw[thick, fill=white] (0,0) circle (0.1cm);
    \draw[very thick] (1,0) -- (1,-2);
    \draw[thick, fill=white] (2*\ds,0) circle (0.1cm);
    \draw[very thick] (2,0) -- (2,-1);
    \draw[thick, fill=white] (4*\ds,0) circle (0.1cm);
	\Vfoldedgate{7*\ds}{-\ds}
	\draw[very thick] (7*\ds-\ds,-\ds+\ds)--(7*\ds-\ds,-\ds+\ds+3*\d);  
	\draw[very thick] (9*\ds-\ds,-\ds+\ds)--(9*\ds-\ds,-\ds+\ds+2*\d);
	\draw[very thick] (5,\d)--(5,-2*\ds);
	\Vfoldedgate{\ds+4*\ds}{-3*\ds}
	\Vfoldedgate{\ds+8*\ds}{-3*\ds}
	\Vfoldedgate{3*\ds}{-\ds-4*\ds}
	\Vfoldedgate{7*\ds}{-\ds-4*\ds}
	\draw[very thick](0,-4*\ds)--(0,-6*\ds);
	\draw[very thick](5,-4*\ds)--(5,-6*\ds);
	\Vfoldedgate{\ds}{-7*\ds}
	\Vfoldedgate{\ds+4*\ds}{-7*\ds}
	\Vfoldedgate{\ds+8*\ds}{-7*\ds}
	\draw[thick, fill=white] (0,-8*\ds) circle (0.1cm);
	%
	%
    \draw[very thick] (0+\dx,0)--(0+\dx,-3);
    \draw[thick, fill=white] (0+\dx,0) circle (0.1cm);
    \draw[very thick] (1+\dx,0) -- (1+\dx,-2);
    \draw[thick, fill=white] (2*\ds+\dx,0) circle (0.1cm);
    \draw[very thick] (2+\dx,0) -- (2+\dx,-1);
    \draw[thick, fill=white] (4*\ds+\dx,0) circle (0.1cm);
	\Vfoldedstargate{7*\ds+\dx}{-\ds}
	\draw[very thick] (7*\ds-\ds+\dx,-\ds+\ds)--(7*\ds-\ds+\dx,-\ds+\ds+3*\d);  
	\draw[very thick] (9*\ds-\ds+\dx,-\ds+\ds)--(9*\ds-\ds+\dx,-\ds+\ds+2*\d);
	\draw[very thick] (5+\dx,\d)--(5+\dx,-2*\ds);
	\Vfoldedstargate{\ds+4*\ds+\dx}{-3*\ds}
	\Vfoldedstargate{\ds+8*\ds+\dx}{-3*\ds}
	\Vfoldedstargate{3*\ds+\dx}{-\ds-4*\ds}
	\Vfoldedstargate{7*\ds+\dx}{-\ds-4*\ds}
	\draw[very thick](0+\dx,-4*\ds)--(0+\dx,-6*\ds);
	\draw[very thick](5+\dx,-4*\ds)--(5+\dx,-6*\ds);
	\Vfoldedstargate{\ds+\dx}{-7*\ds}
	\Vfoldedstargate{\ds+4*\ds+\dx}{-7*\ds}
	\Vfoldedstargate{\ds+8*\ds+\dx}{-7*\ds}
	\draw[thick, fill=white] (0+\dx,-8*\ds) circle (0.1cm);
	%
	%
	\draw[very thick] (7*\ds-\ds,-\ds+\ds+3*\d) -- (7*\ds-\ds+\dx,-\ds+\ds+3*\d);
	\draw[very thick] (9*\ds-\ds,-\ds+\ds+2*\d) -- (9*\ds-\ds+\dx,-\ds+\ds+2*\d);
	\draw[very thick] (11*\ds-\ds,-\ds+\ds+\d) -- (11*\ds-\ds+\dx,-\ds+\ds+\d);
    \end{tikzpicture} =\\
    %
    %
    %
    %
    %
    %
    &=\begin{tikzpicture}[baseline=(current  bounding  box.center), scale=.7]
    \def\ds{0.5}
    \def\d{0.25}
    \def\dx{7}
    \draw[very thick] (0,0)--(0,-3);
    \draw[thick, fill=white] (0,0) circle (0.1cm);
    \draw[very thick] (1,0) -- (1,-2);
    \draw[thick, fill=white] (2*\ds,0) circle (0.1cm);
    \draw[very thick] (2,0) -- (2,-1) -- (3,-1) -- (3,0);
    \draw[thick, fill=white] (4*\ds,0) circle (0.1cm);
	\draw[very thick] (7*\ds-\ds,-\ds+\ds)--(7*\ds-\ds,-\ds+\ds+3*\d);  
	\draw[very thick] (9*\ds-\ds,-\ds+\ds)--(9*\ds-\ds,-\ds+\ds+2*\d);
	\draw[very thick] (5,\d)--(5,-2*\ds);
	\draw[very thick] (4,0) -- (4,-2);
	\Vfoldedgate{7*\ds}{-\ds-4*\ds}
	\draw[very thick](0,-4*\ds)--(0,-6*\ds);
	\draw[very thick](5,0)--(5,-6*\ds);
	\draw[very thick](1,-2)--(3,-2);
	\draw[very thick] (0,-3) -- (2,-3);
	\Vfoldedgate{\ds+4*\ds}{-7*\ds}
	\Vfoldedgate{\ds+8*\ds}{-7*\ds}	
	\draw[very thick,decorate,decoration={bent,aspect=0.3,amplitude=10pt}](0,-4) --(1,-4);
	\draw[thick, fill=white] (0,-8*\ds) circle (0.1cm);
	%
	%
	\draw[very thick] (0+\dx,0)--(0+\dx,-3);
    \draw[thick, fill=white] (0+\dx,0) circle (0.1cm);
    \draw[very thick] (1+\dx,0) -- (1+\dx,-2);
    \draw[thick, fill=white] (2*\ds+\dx,0) circle (0.1cm);
    \draw[very thick] (2+\dx,0) -- (2+\dx,-1) -- (3+\dx,-1) -- (3+\dx,0);
    \draw[thick, fill=white] (4*\ds+\dx,0) circle (0.1cm);
	\draw[very thick] (7*\ds-\ds+\dx,-\ds+\ds)--(7*\ds-\ds+\dx,-\ds+\ds+3*\d);  
	\draw[very thick] (9*\ds-\ds+\dx,-\ds+\ds)--(9*\ds-\ds+\dx,-\ds+\ds+2*\d);
	\draw[very thick] (5+\dx,\d)--(5+\dx,-2*\ds);
	\draw[very thick] (4+\dx,0) -- (4+\dx,-2);
	\Vfoldedstargate{7*\ds+\dx}{-\ds-4*\ds}
	\draw[very thick](0+\dx,-4*\ds)--(0+\dx,-6*\ds);
	\draw[very thick](5+\dx,0)--(5+\dx,-6*\ds);
	\draw[very thick](1+\dx,-2)--(3+\dx,-2);
	\draw[very thick] (0+\dx,-3) -- (2+\dx,-3);
	\Vfoldedstargate{\ds+4*\ds+\dx}{-7*\ds}
	\Vfoldedstargate{\ds+8*\ds+\dx}{-7*\ds}	
	\draw[very thick,decorate,decoration={bent,aspect=0.3,amplitude=10pt}](0+\dx,-4) --(1+\dx,-4);
	\draw[thick, fill=white] (0+\dx,-8*\ds) circle (0.1cm);
	%
	%
	\draw[very thick] (7*\ds-\ds,-\ds+\ds+3*\d) -- (7*\ds-\ds+\dx,-\ds+\ds+3*\d);
	\draw[very thick] (9*\ds-\ds,-\ds+\ds+2*\d) -- (9*\ds-\ds+\dx,-\ds+\ds+2*\d);
	\draw[very thick] (11*\ds-\ds,-\ds+\ds+\d) -- (11*\ds-\ds+\dx,-\ds+\ds+\d);
    \end{tikzpicture},
\end{split}
\end{equation}
where in going to the second line we used \cref{du_1} and in going to the third line we used \cref{du_2} and the identity
\begin{equation}
\begin{tikzpicture}[baseline=(current  bounding  box.center), scale=.7]
\Vfoldedgate{0}{0}
\Vfoldedstargate{2}{0}
\draw[very thick](-0.5,0.5) -- (-0.5,0.5+0.5) -- (1.5,0.5+0.5) -- (1.5,0.5);
\draw[very thick](0.5,0.5) -- (0.5,0.5+0.25) -- (2.5,0.5+0.25) -- (2.5,0.5);
\end{tikzpicture} = 
\begin{tikzpicture}[baseline=(current  bounding  box.center), scale=.7]
\Vfoldedgate{0}{0}
\Vfoldeddaggate{0}{1}
\end{tikzpicture} = 
\begin{tikzpicture}[baseline=(current  bounding  box.center), scale=.7]
%
\draw[very thick] (0,0) -- (0,0.6);
%
\draw[very thick] (0.5,0) -- (0.5,0.6); 
\end{tikzpicture}\, ,
\end{equation} 
that follows from the unitarity of $V$. Similarly,
\begin{equation} \label{pic_2}
\Tr_{11^\prime}(\mathsf{S}_{11^\prime}{\mathcal{U}_{t^{\star}}^\dagger}^{\otimes 2}(\mathsf{S}_{BB^\prime}))=
\begin{tikzpicture}[baseline=(current  bounding  box.center), scale=.7]
    \def\ds{0.5}
    \def\d{0.25}
    \def\dx{7}
    \draw[very thick] (0,0)--(0,-3);
    \draw[thick, fill=white] (0,0) circle (0.1cm);
    \draw[very thick] (1,0) -- (1,-2);
    \draw[thick, fill=white] (2*\ds,0) circle (0.1cm);
    \draw[very thick] (2,0) -- (2,-1);
    \draw[thick, fill=white] (4*\ds,0) circle (0.1cm);
	\draw[very thick] (7*\ds-\ds,-\ds+\ds)--(7*\ds-\ds,-\ds+\ds+3*\d);  
	\draw[very thick] (9*\ds-\ds,-\ds+\ds)--(9*\ds-\ds,-\ds+\ds+2*\d);
	\draw[very thick] (5,\d)--(5,-2*\ds);
	\draw[very thick] (3,0)--(3,-1);
	\draw[very thick] (4,0)--(4,-2);
	\draw[very thick] (5,0)--(5,-3);
	\Vfoldedgate{\ds+4*\ds}{-3*\ds}
	\Vfoldedgate{3*\ds}{-\ds-4*\ds}
	\Vfoldedgate{7*\ds}{-\ds-4*\ds}
	\draw[very thick](0,-4*\ds)--(0,-6*\ds);
	\Vfoldedgate{\ds}{-7*\ds}
	\Vfoldedgate{\ds+4*\ds}{-7*\ds}
	\Vfoldedgate{\ds+8*\ds}{-7*\ds}
	%
	%
    \draw[very thick] (0+\dx,0)--(0+\dx,-3);
    \draw[thick, fill=white] (0+\dx,0) circle (0.1cm);
    \draw[very thick] (1+\dx,0) -- (1+\dx,-2);
    \draw[thick, fill=white] (2*\ds+\dx,0) circle (0.1cm);
    \draw[very thick] (2+\dx,0) -- (2+\dx,-1);
    \draw[thick, fill=white] (4*\ds+\dx,0) circle (0.1cm);
	\draw[very thick] (7*\ds-\ds+\dx,-\ds+\ds)--(7*\ds-\ds+\dx,-\ds+\ds+3*\d);  
	\draw[very thick] (9*\ds-\ds+\dx,-\ds+\ds)--(9*\ds-\ds+\dx,-\ds+\ds+2*\d);
	\draw[very thick] (5+\dx,\d)--(5+\dx,-2*\ds);
	\draw[very thick] (3+\dx,0)--(3+\dx,-1);
	\draw[very thick] (4+\dx,0)--(4+\dx,-2);
	\draw[very thick] (5+\dx,0)--(5+\dx,-3);
	\Vfoldedstargate{\ds+4*\ds+\dx}{-3*\ds}
	\Vfoldedstargate{3*\ds+\dx}{-\ds-4*\ds}
	\Vfoldedstargate{7*\ds+\dx}{-\ds-4*\ds}
	\draw[very thick](0+\dx,-4*\ds)--(0+\dx,-6*\ds);
	\draw[very thick](5+\dx,-4*\ds)--(5+\dx,-6*\ds);
	\Vfoldedstargate{\ds+\dx}{-7*\ds}
	\Vfoldedstargate{\ds+4*\ds+\dx}{-7*\ds}
	\Vfoldedstargate{\ds+8*\ds+\dx}{-7*\ds}
	%
	%
	\draw[very thick] (7*\ds-\ds,-\ds+\ds+3*\d) -- (7*\ds-\ds+\dx,-\ds+\ds+3*\d);
	\draw[very thick] (9*\ds-\ds,-\ds+\ds+2*\d) -- (9*\ds-\ds+\dx,-\ds+\ds+2*\d);
	\draw[very thick] (11*\ds-\ds,-\ds+\ds+\d) -- (11*\ds-\ds+\dx,-\ds+\ds+\d);
	\draw[very thick] (0,-4)--(0,-4-\d)--(\dx,-4-\d)--(\dx,-4);
    \end{tikzpicture}\, .
 \end{equation}
 Using \cref{pic_1,pic_2} we have,
 \begin{equation} \label{res_1}
 \begin{split}
\left\lVert \Tr_{11^\prime}({\mathcal{U}_{t^{\star}}^\dagger}^{\otimes 2}(\mathsf{S}_{BB^\prime})) \right\rVert_2^2&=\begin{tikzpicture}[baseline=(current  bounding  box.center), scale=.7]
    \def\ds{0.5}
    \def\d{0.25}
    \def\dx{7}
    \draw[very thick] (0,0)--(0,-3);
    \draw[thick, fill=white] (0,0) circle (0.1cm);
    \draw[very thick] (1,0) -- (1,-2);
    \draw[thick, fill=white] (2*\ds,0) circle (0.1cm);
    \draw[very thick] (2,0) -- (2,-1) -- (3,-1) -- (3,0);
    \draw[thick, fill=white] (4*\ds,0) circle (0.1cm);
	\draw[very thick] (7*\ds-\ds,-\ds+\ds)--(7*\ds-\ds,-\ds+\ds+3*\d);  
	\draw[very thick] (9*\ds-\ds,-\ds+\ds)--(9*\ds-\ds,-\ds+\ds+2*\d);
	\draw[very thick] (5,\d)--(5,-2*\ds);
	\draw[very thick] (4,0) -- (4,-2);
	\Vfoldedgate{7*\ds}{-\ds-4*\ds}
	\draw[very thick](0,-4*\ds)--(0,-6*\ds);
	\draw[very thick](5,0)--(5,-6*\ds);
	\draw[very thick](1,-2)--(3,-2);
	\draw[very thick] (0,-3) -- (2,-3);
	\Vfoldedgate{\ds+4*\ds}{-7*\ds}
	\Vfoldedgate{\ds+8*\ds}{-7*\ds}	
	\draw[very thick,decorate,decoration={bent,aspect=0.3,amplitude=10pt}](0,-4) --(1,-4);
	\draw[thick, fill=white] (0,-8*\ds) circle (0.1cm);
	%
	%
	\draw[very thick] (0+\dx,0)--(0+\dx,-3);
    \draw[thick, fill=white] (0+\dx,0) circle (0.1cm);
    \draw[very thick] (1+\dx,0) -- (1+\dx,-2);
    \draw[thick, fill=white] (2*\ds+\dx,0) circle (0.1cm);
    \draw[very thick] (2+\dx,0) -- (2+\dx,-1) -- (3+\dx,-1) -- (3+\dx,0);
    \draw[thick, fill=white] (4*\ds+\dx,0) circle (0.1cm);
	\draw[very thick] (7*\ds-\ds+\dx,-\ds+\ds)--(7*\ds-\ds+\dx,-\ds+\ds+3*\d);  
	\draw[very thick] (9*\ds-\ds+\dx,-\ds+\ds)--(9*\ds-\ds+\dx,-\ds+\ds+2*\d);
	\draw[very thick] (5+\dx,\d)--(5+\dx,-2*\ds);
	\draw[very thick] (4+\dx,0) -- (4+\dx,-2);
	\Vfoldedstargate{7*\ds+\dx}{-\ds-4*\ds}
	\draw[very thick](0+\dx,-4*\ds)--(0+\dx,-6*\ds);
	\draw[very thick](5+\dx,0)--(5+\dx,-6*\ds);
	\draw[very thick](1+\dx,-2)--(3+\dx,-2);
	\draw[very thick] (0+\dx,-3) -- (2+\dx,-3);
	\Vfoldedstargate{\ds+4*\ds+\dx}{-7*\ds}
	\Vfoldedstargate{\ds+8*\ds+\dx}{-7*\ds}	
	\draw[very thick,decorate,decoration={bent,aspect=0.3,amplitude=10pt}](0+\dx,-4) --(1+\dx,-4);
	\draw[thick, fill=white] (0+\dx,-8*\ds) circle (0.1cm);
	%
	%
	\draw[very thick] (7*\ds-\ds,-\ds+\ds+3*\d) -- (7*\ds-\ds+\dx,-\ds+\ds+3*\d);
	\draw[very thick] (9*\ds-\ds,-\ds+\ds+2*\d) -- (9*\ds-\ds+\dx,-\ds+\ds+2*\d);
	\draw[very thick] (11*\ds-\ds,-\ds+\ds+\d) -- (11*\ds-\ds+\dx,-\ds+\ds+\d);
    %
    %
    %
    %
    %
	%
	%
	\def\ds{0.5}
    \def\d{0.25}
    \def\dx{7}
    \def\dy{9}
    \draw[very thick] (0,-0-\dy)--(0,3-\dy);
    \draw[thick, fill=white] (0,-0-\dy) circle (0.1cm);
    \draw[very thick] (1,-0-\dy) -- (1,2-\dy);
    \draw[thick, fill=white] (2*\ds,-0-\dy) circle (0.1cm);
    \draw[very thick] (2,-0-\dy) -- (2,1-\dy) -- (3,1-\dy) -- (3,-0-\dy);
    \draw[thick, fill=white] (4*\ds,-0-\dy) circle (0.1cm);
	\draw[very thick] (7*\ds-\ds,\ds-\ds-\dy)--(7*\ds-\ds,\ds-\ds-3*\d-\dy);  
	\draw[very thick] (9*\ds-\ds,\ds-\ds-\dy)--(9*\ds-\ds,\ds-\ds-2*\d-\dy);
	\draw[very thick] (5,-\d-\dy)--(5,2*\ds-\dy);
	\draw[very thick] (4,0-\dy) -- (4,2-\dy);
	\Vfoldeddaggate{7*\ds}{\ds+4*\ds-\dy}
	\draw[very thick](0,4*\ds-\dy)--(0,6*\ds-\dy);
	\draw[very thick](5,0-\dy)--(5,6*\ds-\dy);
	\draw[very thick](1,2-\dy)--(3,2-\dy);
	\draw[very thick] (0,3-\dy) -- (2,3-\dy);
	\Vfoldeddaggate{\ds+4*\ds}{7*\ds-\dy}
	\Vfoldeddaggate{\ds+8*\ds}{7*\ds-\dy}	
	\draw[very thick,decorate,decoration={bent,aspect=0.3,amplitude=-10pt}](0,4-\dy) --(1,4-\dy);
	\draw[thick, fill=white] (0,8*\ds-\dy) circle (0.1cm);
	%
	%
	%
    \draw[very thick] (0+\dx,-0-\dy)--(0+\dx,3-\dy);
    \draw[thick, fill=white] (0+\dx,-0-\dy) circle (0.1cm);
    \draw[very thick] (1+\dx,-0-\dy) -- (1+\dx,2-\dy);
    \draw[thick, fill=white] (2*\ds+\dx,-0-\dy) circle (0.1cm);
    \draw[very thick] (2+\dx,-0-\dy) -- (2+\dx,1-\dy) -- (3+\dx,1-\dy) -- (3+\dx,-0-\dy);
    \draw[thick, fill=white] (4*\ds+\dx,-0-\dy) circle (0.1cm);
	\draw[very thick] (7*\ds-\ds+\dx,\ds-\ds-\dy)--(7*\ds-\ds+\dx,\ds-\ds-3*\d-\dy);  
	\draw[very thick] (9*\ds-\ds+\dx,\ds-\ds-\dy)--(9*\ds-\ds+\dx,\ds-\ds-2*\d-\dy);
	\draw[very thick] (5+\dx,-\d-\dy)--(5+\dx,2*\ds-\dy);
	\draw[very thick] (4+\dx,0-\dy) -- (4+\dx,2-\dy);
	\Vfoldedstardaggate{7*\ds+\dx}{\ds+4*\ds-\dy}
	\draw[very thick](0+\dx,4*\ds-\dy)--(0+\dx,6*\ds-\dy);
	\draw[very thick](5+\dx,0-\dy)--(5+\dx,6*\ds-\dy);
	\draw[very thick](1+\dx,2-\dy)--(3+\dx,2-\dy);
	\draw[very thick] (0+\dx,3-\dy) -- (2+\dx,3-\dy);
	\Vfoldedstardaggate{\ds+4*\ds+\dx}{7*\ds-\dy}
	\Vfoldedstardaggate{\ds+8*\ds+\dx}{7*\ds-\dy}	
	\draw[very thick,decorate,decoration={bent,aspect=0.3,amplitude=-10pt}](0+\dx,4-\dy) --(1+\dx,4-\dy);
	\draw[thick, fill=white] (0+\dx,8*\ds-\dy) circle (0.1cm);
	\draw[very thick] (6*\ds,-3*\d-\dy) -- (6*\ds+\dx,-3*\d-\dy);
	\draw[very thick] (8*\ds,-2*\d-\dy) -- (8*\ds+\dx,-2*\d-\dy);
	\draw[very thick] (10*\ds,-\d-\dy) -- (10*\ds+\dx,-\d-\dy);
	%
	%
	%
	\draw[very thick](1,-4)--(1,-5);
	\draw[very thick](2,-4)--(2,-5);
	\draw[very thick](3,-4)--(3,-5);
	\draw[very thick](4,-4)--(4,-5);
	\draw[very thick](5,-4)--(5,-5);
	\draw[very thick](1+\dx,-4)--(1+\dx,-5);
	\draw[very thick](2+\dx,-4)--(2+\dx,-5);
	\draw[very thick](3+\dx,-4)--(3+\dx,-5);
	\draw[very thick](4+\dx,-4)--(4+\dx,-5);
	\draw[very thick](5+\dx,-4)--(5+\dx,-5);
    \end{tikzpicture} =\\
    %
    %
    &={\begin{tikzpicture}[baseline=(current  bounding  box.center), scale=.7]
    \draw[very thick] (0,0) -- (0,-1);
    \draw[thick, fill=white] (0,0) circle (0.1cm);
    \draw[thick, fill=white] (0,-1) circle (0.1cm);
    \end{tikzpicture}}^{\, 6+2} \quad
    {\begin{tikzpicture}[baseline=(current  bounding  box.center), scale=.7]
    \draw[very thick] (0,0) rectangle (1,-1);
    \end{tikzpicture}}^{\, 3-1}=
    {\begin{tikzpicture}[baseline=(current  bounding  box.center), scale=.7]
    \draw[very thick] (0,0) -- (0,-1);
    \draw[thick, fill=white] (0,0) circle (0.1cm);
    \draw[thick, fill=white] (0,-1) circle (0.1cm);
    \end{tikzpicture}}^{\, L+2} \quad
    {\begin{tikzpicture}[baseline=(current  bounding  box.center), scale=.7]
    \draw[very thick] (0,0) rectangle (1,-1);
    \end{tikzpicture}}^{\, \frac{L}{2}-1} = q^{2L},
\end{split}
\end{equation}
where we used that
\begin{equation}
{\begin{tikzpicture}[baseline=(current  bounding  box.center), scale=.7]
    \draw[very thick] (0,0) -- (0,-1);
    \draw[thick, fill=white] (0,0) circle (0.1cm);
    \draw[thick, fill=white] (0,-1) circle (0.1cm);
    \end{tikzpicture}}=q, \quad
    {\begin{tikzpicture}[baseline=(current  bounding  box.center), scale=.7]
    \draw[very thick] (0,0) rectangle (1,-1);
    \end{tikzpicture}}=q^2.
\end{equation}
Similarly,
\begin{equation} \label{res_2}
\begin{split}
\left\langle \Tr_{11^\prime}({\mathcal{U}_{t^{\star}}^\dagger}^{\otimes 2}(\mathsf{S}_{BB^\prime})), \Tr_{11^\prime}(\mathsf{S}_{11^\prime}{\mathcal{U}_{t^{\star}}^\dagger}^{\otimes 2}(\mathsf{S}_{BB^\prime})) \right\rangle&=
\begin{tikzpicture}[baseline=(current  bounding  box.center), scale=.7]
    \draw[very thick] (0,0) -- (0,-1);
    \draw[thick, fill=white] (0,0) circle (0.1cm);
    \draw[thick, fill=white] (0,-1) circle (0.1cm);
\end{tikzpicture} \quad
{\begin{tikzpicture}[baseline=(current  bounding  box.center), scale=.7]
 \draw[very thick] (0,0) rectangle (1,-1);
\end{tikzpicture}}^{\, 3-1} \quad
\begin{tikzpicture}[baseline=(current  bounding  box.center), scale=.7]
    \def\ds{0.5}
    \def\d{0.25}
    \def\dx{7}
    \draw[very thick] (0,0)--(0,-3);
    \draw[thick, fill=white] (0,0) circle (0.1cm);
    \draw[very thick] (1,0) -- (1,-2);
    \draw[thick, fill=white] (2*\ds,0) circle (0.1cm);
    \draw[very thick] (2,0) -- (2,-1);
    \draw[thick, fill=white] (4*\ds,0) circle (0.1cm);
	\draw[very thick] (7*\ds-\ds,-\ds+\ds)--(7*\ds-\ds,-\ds+\ds+3*\d);  
	\draw[very thick] (3,0)--(3,-1);
	\Vfoldedgate{\ds+4*\ds}{-3*\ds}
	\draw[thick , fill=white](\ds+4*\ds+0.5,-3*\ds-0.5) circle (0.1cm);
	\Vfoldedgate{3*\ds}{-\ds-4*\ds}
	\draw[thick , fill=white](3*\ds+0.5,-5*\ds-0.5) circle (0.1cm);
	\draw[very thick](0,-4*\ds)--(0,-6*\ds);
	\Vfoldedgate{\ds}{-7*\ds}
	\draw[thick , fill=white](\ds+0.5,-7*\ds-0.5) circle (0.1cm);
	%
	%
    \draw[very thick] (0+\dx,0)--(0+\dx,-3);
    \draw[thick, fill=white] (0+\dx,0) circle (0.1cm);
    \draw[very thick] (1+\dx,0) -- (1+\dx,-2);
    \draw[thick, fill=white] (2*\ds+\dx,0) circle (0.1cm);
    \draw[very thick] (2+\dx,0) -- (2+\dx,-1);
    \draw[thick, fill=white] (4*\ds+\dx,0) circle (0.1cm);
	\draw[very thick] (7*\ds-\ds+\dx,-\ds+\ds)--(7*\ds-\ds+\dx,-\ds+\ds+3*\d);  
	\draw[very thick] (3+\dx,0)--(3+\dx,-1);
	\Vfoldedstargate{\ds+4*\ds+\dx}{-3*\ds}
	\draw[thick , fill=white](\ds+4*\ds+\dx+0.5,-3*\ds-0.5) circle (0.1cm);
	\Vfoldedstargate{3*\ds+\dx}{-\ds-4*\ds}
	\draw[thick , fill=white](3*\ds+\dx+0.5,-5*\ds-0.5) circle (0.1cm);
	\draw[very thick](0+\dx,-4*\ds)--(0+\dx,-6*\ds);
	\Vfoldedstargate{\ds+\dx}{-7*\ds}
	\draw[thick , fill=white](\ds+\dx+0.5,-7*\ds-0.5) circle (0.1cm);
	%
	%
	\draw[very thick] (7*\ds-\ds,-\ds+\ds+3*\d) -- (7*\ds-\ds+\dx,-\ds+\ds+3*\d);
	\draw[very thick] (0,-4)--(0,-4-\d)--(\dx,-4-\d)--(\dx,-4);
    \end{tikzpicture}=\\
    &=q^{2L-1} \lVert M_-^{\frac{L}{2}} \rVert_2^2,
 \end{split}
 \end{equation}
where $M_-$ is defined in \cref{matrices}. Notice that $M_-$ is a $4x4$ matrix where the input and output is still defined from bottom to top, namely in the light-cone direction. Finally,
\begin{equation} \label{res_3}
\begin{split}
\left\lVert \Tr_{11^\prime}(\mathsf{S}_{11^\prime}{\mathcal{U}_{t^{\star}}^\dagger}^{\otimes 2}(\mathsf{S}_{BB^\prime})) \right\rVert_2^2&=
{\begin{tikzpicture}[baseline=(current  bounding  box.center), scale=.7]
 \draw[very thick] (0,0) rectangle (1,-1);
\end{tikzpicture}}^{\, 3-1} \quad
\begin{tikzpicture}[baseline=(current  bounding  box.center), scale=.7]
    \def\ds{0.5}
    \def\d{0.25}
    \def\dx{7}
    \def\dy{9}
    \draw[very thick] (0,0)--(0,-3);
    \draw[thick, fill=white] (0,0) circle (0.1cm);
    \draw[very thick] (1,0) -- (1,-2);
    \draw[thick, fill=white] (2*\ds,0) circle (0.1cm);
    \draw[very thick] (2,0) -- (2,-1);
    \draw[thick, fill=white] (4*\ds,0) circle (0.1cm);
	\draw[very thick] (7*\ds-\ds,-\ds+\ds)--(7*\ds-\ds,-\ds+\ds+3*\d);  
	\draw[very thick] (3,0)--(3,-1);
	\Vfoldedgate{\ds+4*\ds}{-3*\ds}
	\Vfoldedgate{3*\ds}{-\ds-4*\ds}
	\draw[very thick](0,-4*\ds)--(0,-6*\ds);
	\Vfoldedgate{\ds}{-7*\ds}
	%
	%
    \draw[very thick] (0+\dx,0)--(0+\dx,-3);
    \draw[thick, fill=white] (0+\dx,0) circle (0.1cm);
    \draw[very thick] (1+\dx,0) -- (1+\dx,-2);
    \draw[thick, fill=white] (2*\ds+\dx,0) circle (0.1cm);
    \draw[very thick] (2+\dx,0) -- (2+\dx,-1);
    \draw[thick, fill=white] (4*\ds+\dx,0) circle (0.1cm);
	\draw[very thick] (7*\ds-\ds+\dx,-\ds+\ds)--(7*\ds-\ds+\dx,-\ds+\ds+3*\d);  
	\draw[very thick] (3+\dx,0)--(3+\dx,-1);
	\Vfoldedstargate{\ds+4*\ds+\dx}{-3*\ds}
	\Vfoldedstargate{3*\ds+\dx}{-\ds-4*\ds}
	\draw[very thick](0+\dx,-4*\ds)--(0+\dx,-6*\ds);
	\Vfoldedstargate{\ds+\dx}{-7*\ds}
	%
	%
	\draw[very thick] (7*\ds-\ds,-\ds+\ds+3*\d) -- (7*\ds-\ds+\dx,-\ds+\ds+3*\d);
	\draw[very thick] (0,-4)--(0,-4-\d)--(\dx,-4-\d)--(\dx,-4);
	%
	%
	%
	%
	\draw[very thick] (0,0-\dy)--(0,3-\dy);
    \draw[thick, fill=white] (0,0-\dy) circle (0.1cm);
    \draw[very thick] (1,0-\dy) -- (1,2-\dy);
    \draw[thick, fill=white] (2*\ds,0-\dy) circle (0.1cm);
    \draw[very thick] (2,0-\dy) -- (2,1-\dy);
    \draw[thick, fill=white] (4*\ds,0-\dy) circle (0.1cm);
	\draw[very thick] (7*\ds-\ds,-\dy)--(7*\ds-\ds,-3*\d-\dy);  
	\draw[very thick] (3,0-\dy)--(3,1-\dy);
	\Vfoldeddaggate{\ds+4*\ds}{3*\ds-\dy}
	\Vfoldeddaggate{3*\ds}{5*\ds-\dy}
	\draw[very thick](0,4*\ds-\dy)--(0,6*\ds-\dy);
	\Vfoldeddaggate{\ds}{7*\ds-\dy}
	%
	%
    \draw[very thick] (0+\dx,0-\dy)--(0+\dx,3-\dy);
    \draw[thick, fill=white] (0+\dx,0-\dy) circle (0.1cm);
    \draw[very thick] (1+\dx,0-\dy) -- (1+\dx,2-\dy);
    \draw[thick, fill=white] (2*\ds+\dx,0-\dy) circle (0.1cm);
    \draw[very thick] (2+\dx,0-\dy) -- (2+\dx,1-\dy);
    \draw[thick, fill=white] (4*\ds+\dx,0-\dy) circle (0.1cm);
	\draw[very thick] (7*\ds-\ds+\dx,-\dy)--(7*\ds-\ds+\dx,-3*\d-\dy);  
	\draw[very thick] (3+\dx,0-\dy)--(3+\dx,1-\dy);
	\Vfoldedstardaggate{\ds+4*\ds+\dx}{3*\ds-\dy}
	\Vfoldedstardaggate{3*\ds+\dx}{5*\ds-\dy}
	\draw[very thick](0+\dx,4*\ds-\dy)--(0+\dx,6*\ds-\dy);
	\Vfoldedstardaggate{\ds+\dx}{7*\ds-\dy}
	%
	%
	\draw[very thick] (7*\ds-\ds,-3*\d-\dy) -- (7*\ds-\ds+\dx,-3*\d-\dy);
	\draw[very thick] (0,4-\dy)--(0,4+\d-\dy)--(\dx,4+\d-\dy)--(\dx,4-\dy);
	\draw[very thick] (1,-4) -- (1,-5);
	\draw[very thick] (2,-3) -- (2,-6);
	\draw[very thick] (3,-2) -- (3,-7);
	\draw[very thick] (1+\dx,-4) -- (1+\dx,-5);
	\draw[very thick] (2+\dx,-3) -- (2+\dx,-6);
	\draw[very thick] (3+\dx,-2) -- (3+\dx,-7);
    \end{tikzpicture} = \\
    &= q^{2L-2} \lVert P^{\frac{L}{2}} \rVert_2^2,
    \end{split}
    \end{equation}
where
\begin{equation}
P \coloneqq \frac{1}{q}
\begin{tikzpicture}[baseline=(current  bounding  box.center), scale=.7]
\def\d{0.2}
\Vfoldedgate{0}{0}
\draw[thick, fill=white] (-0.5,0.5) circle (0.1cm);
\Vfoldeddaggate{0}{-1.5}
\draw[thick, fill=white] (-0.5,-1.5-0.5) circle (0.1cm);
\draw[very thick] (0.5,-0.5) -- (+0.5,-1);
\node at (-0.5-\d,-0.5) {$i_1$};
\node at (-0.5-\d,-1) {$i_2$};
\node at (0.5+\d+0.1,0.5) {$o_1$};
\node at (0.5+\d+0.1,-2) {$o_2$};
\end{tikzpicture}=
%
%
\frac{1}{q}
\begin{tikzpicture}[baseline=(current  bounding  box.center), scale=.7]
\def\d{0.2}
\draw[very thick] (-0.5, +0.5) -- (+0.25,-0.25);
\draw[very thick] (-0.5,-0.5) -- (+0.5,+0.5);
\draw[ thick, fill=myred, rounded corners=2pt] (-0.25,+0.25) rectangle (+0.25,-0.25);
\draw[thick] (0,+0.15) -- (+0.15,+0.15) -- (+0.15,0);
\def\dx{2}
\draw[very thick] (\dx-0.5, +0.5) -- (\dx+0.25,-0.25);
\draw[very thick] (\dx-0.5,-0.5) -- (\dx+0.5,+0.5);
\draw[ thick, fill=myblue, rounded corners=2pt] (\dx-0.25,+0.25) rectangle (\dx+0.25,-0.25);
\draw[thick] (\dx,+0.15) -- (\dx+0.15,+0.15) -- (\dx+0.15,0);
\draw[thick, fill=white] (-0.5,0.5) circle (0.1cm);  
\draw[thick, fill=white] (\dx-0.5,0.5) circle (0.1cm);
\draw[very thick,decorate,decoration={bent,aspect=0.3,amplitude=-20pt}] (0.25,-0.25) -- (\dx+0.25,-0.25);
\node at (-0.5-\d,-0.5) {$i_1$};
\node at (\dx-0.5-\d,-0.5) {$i_2$};
\node at (0.5+\d+0.1,0.5) {$o_1$};
\node at (0.5+\dx+\d+0.1,0.5) {$o_2$};
\end{tikzpicture},
\end{equation}
where we used the indices $i_{1,2}$, $o_{1,2}$ to denote the input and output spaces. Using \cref{res_1,res_2,res_3} in \cref{opentloc_proof}, we get \cref{locopent}. Note that the properties of $M_+,M_-$ in \cref{property_1,property_3} have been established in previous works \cite{bertiniExactCorrelationFunctions2019,aravindaDualunitaryQuantumBernoulli2021}, while \cref{property_2} follows by some graphical manipulations,
\begin{equation}
\begin{split}
\lVert P^2 \rVert &= \frac{1}{q^2}
\begin{tikzpicture}[baseline=(current  bounding  box.center), scale=.7]
\def\ds{0.25}
\Vfoldedgate{0}{0}
\draw[thick, fill=white] (0-0.5,0+0.5) circle (0.1cm);
\Vfoldedstargate{1.5}{0}
\draw[thick, fill=white] (1.5-0.5,0+0.5) circle (0.1cm);
\Vfoldedstargate{3}{0}
\draw[thick, fill=white] (3-0.5,0+0.5) circle (0.1cm);
\Vfoldedgate{4.5}{0}
\draw[thick, fill=white] (4.5-0.5,0+0.5) circle (0.1cm);
\draw[very thick] (-0.5,-0.5) -- (-0.5,-0.5-4*\ds) -- (2.5,-0.5-4*\ds) -- (2.5,-0.5);
\draw[very thick] (0.5,-0.5) -- (0.5,-0.5-\ds) -- (2,-0.5-\ds) -- (2,-0.5);
\draw[very thick] (1,-0.5) -- (1,-0.5-3*\ds) -- (4,-0.5-3*\ds) -- (4,-0.5);
\draw[very thick] (3.5,-0.5) -- (3.5,-0.5-2*\ds) -- (5,-0.5-2*\ds) -- (5,-0.5);
\draw[very thick] (0.5,0.5) -- (0.5,0.5+2*\ds) -- (3.5,0.5+2*\ds) -- (3.5,0.5);
\draw[very thick] (2,0.5) -- (2,0.5+\ds) -- (5,0.5+\ds) -- (5,0.5);
\end{tikzpicture} =
\frac{1}{q^2}
\begin{tikzpicture}[baseline=(current  bounding  box.center), scale=.7]
\def\ds{2}
\def\d{0.25}
\Vfoldedstargate{0}{0}
\draw[thick, fill=white] (0-0.5,0-0.5) circle (0.1cm);
\draw[thick, fill=white] (0+0.5,0+0.5) circle (0.1cm);
\Vfoldedgate{\ds}{0}
\draw[thick, fill=white] (\ds-0.5,0-0.5) circle (0.1cm);
\draw[thick, fill=white] (\ds+0.5,0+0.5) circle (0.1cm);
\Vfoldedgate{0}{-\ds}
\draw[thick, fill=white] (0-0.5,-\ds-0.5) circle (0.1cm);
\draw[thick, fill=white] (0+0.5,-\ds+0.5) circle (0.1cm);
\Vfoldedstargate{\ds}{-\ds}
\draw[thick, fill=white] (\ds-0.5,-\ds-0.5) circle (0.1cm);
\draw[thick, fill=white] (\ds+0.5,-\ds+0.5) circle (0.1cm);
\draw[very thick] (-0.5,0.5) -- (-0.5-\d,0.5) -- (-0.5-\d,-\ds+0.5) -- (-0.5,-\ds+0.5);
\draw[very thick] (-0.5+\ds,0.5) -- (-0.5-\d+\ds,0.5) -- (-0.5-\d+\ds,-\ds+0.5) -- (-0.5+\ds,-\ds+0.5);
\draw[very thick] (0.5,-0.5) -- (0.5,-0.5-\d) -- (0.5+\ds,-0.5-\d) -- (0.5+\ds,-0.5);
\draw[very thick] (0.5,-0.5-\ds) -- (0.5,-0.5-\d-\ds) -- (0.5+\ds,-0.5-\d-\ds) -- (0.5+\ds,-0.5-\ds);
\end{tikzpicture} =\\
&= \frac{1}{q^2} \left\lVert
\begin{tikzpicture}[baseline=(current  bounding  box.center), scale=.7]
\def\ds{2}
\def\d{0.25}
\Vfoldedgate{0}{0}
\draw[thick, fill=white] (0-0.5,0-0.5) circle (0.1cm);
\draw[thick, fill=white] (0+0.5,0+0.5) circle (0.1cm);
\Vfoldeddaggate{0}{\ds}
\draw[thick, fill=white] (-0.5,\ds+0+0.5) circle (0.1cm);
\draw[thick, fill=white] (0.5,\ds+0-0.5) circle (0.1cm);
\draw[very thick](-0.5,0.5)--(-0.5,-0.5+\ds);
\end{tikzpicture}
\right\rVert_2^2 = q^2 \left\lVert M_+ M_+^\dagger \right\rVert_2^2,
\end{split}
\end{equation}
where the first line involves ``unfolding'' the diagram and then ``folding'' it in a different way.

\section{Gradient ascent algorithm for \cref{subsec:bound}} \label{appendix_gradient}
The gradient ascent algorithm is due to Abrudan et al. \cite{abrudanSteepestDescentAlgorithms2008}, which we have previously used in Ref. \cite{andreadakisLongtimeQuantumScrambling2024a}, see Appendix 4 therein. The main idea is that the direction of steepest ascent of $E_p(U)$ at a point $U$ in the space of unitaries $U(d)$ is given as $G_U \coloneqq \Gamma_U U^\dagger - U \Gamma_U^\dagger$, where $\Gamma_U \coloneqq \nabla_{\overline{U}} E_p(\mathcal{U})$ is the gradient on an ambient Euclidean space $\mathbb{R}^{2d \times 2d}$ \cite{abrudanSteepestDescentAlgorithms2008} with $\overline{U}$ the complex conjugate of $U$. For a function $f(U,\overline{U}):\mathbb{R}^{2d \times 2d} \rightarrow \mathbb{R}$, $\nabla_{\overline{U}} f$ is defined as
\cite{abrudanSteepestDescentAlgorithms2008}
\begin{equation}
\nabla_{\overline{U}}f(U,\overline{U}) = \frac{1}{2} \left( \frac{\partial E_p}{\partial U_R} + i \frac{\partial E_p}{\partial U_I} \right),
\end{equation}
where $U_R$ and $U_I$ denote the real and imaginary part of $U$.
Using \cref{opentp_sym} we have
\begin{equation}
\nabla_{\overline{U}}E_p(\mathcal{U}) = \frac{2}{E(\mathsf{S})} \left( \left(1- \frac{E(U)}{E(\mathsf{S})} -\frac{1}{d} \frac{E(U\mathsf{S})}{E(\mathsf{S})} \right) \nabla_{\overline{U}} E(U) + \left(1- \frac{E(U\mathsf{S})}{E(\mathsf{S})} -\frac{1}{d} \frac{E(U)}{E(\mathsf{S})} \right) \nabla_{\overline{U}} E(U\mathsf{S}) \right).
\end{equation}
Also, for a real-valued function $f(U,\overline{U})$, we have $\delta f = \frac{\partial f}{\partial U_R} \delta U_R + \frac{\partial f}{\partial U_I} \delta U_I= 2 \Re\left( \Tr[\left( \nabla_{\overline{U}} f \right)^\dagger \delta U ]\right)$. Performing the variation for $E(U)$ and $E(U\mathsf{S})$ using \cref{opent}, we get
\begin{equation}
\begin{split}
&\nabla_{\overline{U}} E(U) = -\frac{2}{d^2} \Tr_{A^\prime B^\prime}((\mathds{1} \otimes U^\dagger) \mathsf{S}_{AA^\prime} U\otimes U \mathsf{S}_{AA^\prime}) \\
&\nabla_{\overline{U}} E(U\mathsf{S}) = -\frac{2}{d^2} \Tr_{A^\prime B^\prime}((\mathds{1} \otimes U^\dagger) \mathsf{S}_{BB^\prime} U\otimes U \mathsf{S}_{AA^\prime}).
\end{split}
\end{equation}
Using the above equations, we can compute $G_U$ at every point $U$. Then, starting from a random unitary $U_0$, we iteratively update the unitary as $U_{i+1} = \exp(\mu \, \Gamma_{U_i})\,  U_i$, where the step size $\mu$ is dynamically adjusted to enhance the algorithm efficiency \cite{abrudanSteepestDescentAlgorithms2008}. The algorithm terminates when the convergence condition $\lvert E_p(\mathcal{U}_{i+1}) - E_p(\mathcal{U}_i) \rvert \leq \epsilon$ is met. We set $\epsilon \coloneqq 10^{-16}$ which is the order of the machine precision for float64 numbers.

\section{Gate choices for \cref{subsec:DU}} \label{appendix_du}
The $SU(2)$ matrices $v_+,v_-$ in \cref{du_char} can be parametrized as
\begin{align}
v=\begin{pmatrix}
 r e^{i \omega/2} 				& - \sqrt{1-r^2} e^{-i \theta/2} \\
 \sqrt{1-r^2} e^{i \theta/2}	& r e^{- i \omega/2}  
 \end{pmatrix},
 \qquad r\in[0,1],\quad\omega,\theta\in[0,4\pi].
\end{align}
Following Ref. \cite{bertiniOperatorEntanglementLocal2022}, we choose $v_+=v_-$ and $r=0.5$, $\omega = 0.7$, $\theta =0$ for the numerical simulations reported in \cref{subsec:DU}.

\end{document}